\documentclass[11pt,twocolumn]{article}

\usepackage[utf8]{inputenc}
\usepackage[T1]{fontenc}
\usepackage{amsmath,amssymb,amsfonts}
\usepackage{graphicx}
\usepackage{booktabs}
\usepackage{array}
\usepackage{tabularx}
\usepackage{multirow}
\usepackage{xcolor}
\usepackage{listings}
\usepackage{url}
\usepackage{hyperref}
\usepackage{cleveref}
\usepackage[margin=0.85in]{geometry}
\usepackage{enumitem}
\usepackage{caption}
\usepackage{subcaption}
\usepackage{fancyhdr}
\usepackage{balance}
\usepackage{float}

\definecolor{keyword}{RGB}{0,0,180}
\definecolor{comment}{RGB}{100,100,100}
\definecolor{string}{RGB}{160,32,32}
\definecolor{bg}{RGB}{248,248,248}

\lstdefinelanguage{Arch}{
  morekeywords={module,end,pipeline,stage,fsm,state,fifo,arbiter,
    param,port,in,out,inout,const,type,
    comb,seq,on,rising,falling,reg,let,wire,
    if,else,match,case,default,for,
    generate_for,generate_if,generate_else,
    assert,cover,assume,
    struct,enum,interface,domain,inst,package,use,bus,
    handshake,send,receive,
    thread,wait,until,fork,join,once,lock,resource,mutex,cycle,
    synchronizer,kind,latch,clkgate,
    UInt,SInt,Bool,Bit,Clock,Reset,Vec,Token,
    true,false,todo,
    stall,flush,when,
    hook,template,implements,function,return},
  sensitive=true,
  morecomment=[l]{//},
  morestring=[b]",
  morestring=[b]',
}

\lstdefinelanguage{SystemVerilog}{
  morekeywords={module,endmodule,always_ff,always_comb,assign,
    input,output,logic,wire,reg,parameter,localparam,
    if,else,case,endcase,begin,end,posedge,negedge,
    generate,endgenerate,genvar,for,import},
  sensitive=true,
  morecomment=[l]{//},
  morestring=[b]",
}

\setlist{nosep,leftmargin=1.2em}

\title{\textbf{Arch: An AI-Native Hardware Description Language\\for Register-Transfer Clocked Hardware Design}}

\author{
  Shuqing Zhao \\
  \texttt{arch.hdl.lang@gmail.com}
}

\date{April 2026}

\begin{document}
\maketitle

\begin{abstract}
We present \textbf{Arch} (AI-native Register-transfer Clocked Hardware), a hardware description language designed from first principles for micro-architecture specification and AI-assisted code generation. Arch provides first-class constructs for pipelines, FSMs, FIFOs, arbiters, register files, buses, clock-domain crossings, and multi-cycle threads---structures that existing HDLs express only as user-defined patterns prone to subtle errors.

A central design choice is that clocks and resets are parameterized types (\texttt{Clock<D>}, \texttt{Reset<S,P,D?>}) rather than ordinary nets, converting clock-domain crossing (CDC) and reset-domain crossing (RDC) analysis from external linter passes into compile-time typing rules. Combined with tracking of bit widths, port directions, single-driver ownership, and combinational acyclicity, the type system catches multiple drivers, undriven ports, implicit latches, width mismatches, combinational loops, and unsynchronized crossings before any simulator runs. A \texttt{guard} clause on \texttt{reg} declarations captures the valid-data pattern declaratively, catching the producer-side bug where a valid flag asserts before data is written.

Every syntactic choice is governed by an \emph{AI-generatability contract}: an LL(1) grammar, no preprocessor, a uniform declaration schema, named block endings, directional connect arrows, and a \texttt{todo!}\ escape hatch let LLMs produce structurally correct, type-safe Arch from natural-language specs without fine-tuning.

The compiler emits deterministic, lint-clean IEEE~1800-2017 SystemVerilog and auto-generates safety properties (FIFO no-overflow, counter range, FSM legal-state) verified with EBMC. An integrated simulator compiles designs to native C++ and supports both C++ and Python testbenches via a cocotb-compatible API. Case studies cover an L1 data cache and an AXI DMA controller (Yosys/OpenSTA, Sky130); 441 of 443 in-scope VerilogEval and CVDP tests pass, with the 2 remaining failures attributable to reference-testbench bugs rather than Arch-generated designs.
\end{abstract}

\section{Introduction}\label{sec:intro}

The hardware description language landscape is dominated by two IEEE standards---Verilog/SystemVerilog~\cite{ieee1800} and VHDL~\cite{ieee1076}---designed in the 1980s and 1990s, well before modern type theory, formal verification integration, or large language model (LLM)--assisted code generation became practical.
While SystemVerilog has grown to encompass verification constructs (UVM), constrained randomization, and assertions (SVA), the core RTL subset retains characteristics that produce hardware bugs detectable only at simulation time:
implicit latches from incomplete \texttt{if} coverage, multiply-driven nets from accidental concurrent assignments, unsynchronized clock-domain crossings from untracked clock tags, and width mismatches from implicit truncation and extension rules.

Modern alternatives---Chisel~\cite{chisel}, SpinalHDL~\cite{spinalhdl}, Amaranth~\cite{amaranth}, and PyRTL~\cite{pyrtl}---embed RTL in host languages (Scala, Python) and provide varying degrees of parameterization and type safety.
However, embedding inherits host-language complexity (JVM runtime, Scala implicits, Python dynamic typing), and micro-architectural constructs such as pipelines with hazard detection, arbiters with fairness policies, and clock-domain crossings remain user-defined library patterns, not compiler-verified first-class constructs.

Simultaneously, LLMs have demonstrated remarkable ability to generate code from natural-language descriptions~\cite{codex,copilot}.
Yet hardware generation remains brittle: Verilog's irregular syntax, context-dependent semantics, and invisible failure modes (e.g., inferred latches) make LLM-generated HDL unreliable without extensive post-generation verification.
No existing HDL was designed with AI generatability as a first-class design constraint.

A key insight motivating Arch is that \emph{the critical metric for AI-assisted hardware design is iteration latency}---the time from generating a candidate design to discovering whether it is correct. In traditional HDL workflows, many categories of bugs (latches, multiply-driven nets, width mismatches, CDC violations) are invisible until simulation, and simulation itself is slow: commercial event-driven simulators carry heavyweight runtimes because the language permits constructs (delta cycles, X/Z propagation, non-deterministic \texttt{always} ordering) that require dynamic resolution at every time step. An AI agent iterating on a hardware design pays this latency cost on every cycle of generate-compile-simulate. Shortening this iteration radius requires two complementary strategies: (1)~moving error detection from simulation time to compile time, so that the agent receives immediate structured feedback without running a simulator, and (2)~accelerating simulation itself, so that the bugs that \emph{can} only be found at runtime are found faster.

Arch addresses these gaps simultaneously. Its contributions are:

\begin{enumerate}
\item \textbf{First-class micro-architecture constructs.} Pipeline, FSM, FIFO, arbiter, register file, synchronizer, and bus (with a \texttt{handshake} sub-construct for valid/ready/req-ack channels) are language keywords with compiler-verified semantics, not library patterns. Hooks allow customizable behavior injection, and templates enforce structural contracts across modules (\Cref{sec:constructs}).

\item \textbf{Four-dimensional static type system with first-class clock and reset.} Bit widths, clock domains, port directions, and signal ownership are tracked simultaneously, and clocks and resets are themselves parameterized types (\texttt{Clock<D>}, \texttt{Reset<S,P,D?>}) rather than ordinary 1-bit nets. This converts CDC and RDC analysis from external linter passes into compile-time typing rules. Every mismatch is a compile-time error (\Cref{sec:types}).

\item \textbf{AI-generatability contract.} A uniform declaration schema, an LL(1) grammar, named block endings, directional connect arrows, and a \texttt{todo!}\ placeholder enable LLMs to produce correct Arch without fine-tuning (\Cref{sec:ai}).

\item \textbf{Deterministic, lint-clean output.} The compiler emits one SystemVerilog file per top-level module with no latches, no X-propagation, no multiply-driven nets, and no implicit CDCs (\Cref{sec:compiler}).

\item \textbf{Integrated simulation toolchain.} The \texttt{arch sim} command independently generates compiled C++ models for cycle-accurate simulation with assertion evaluation, waveform output, and CDC randomization. Verilator~\cite{verilator} serves separately as the verification oracle for generated SystemVerilog correctness during compiler development. Language-level invariants (no combinational loops, single-driver rule, static clock schedules) further enable a future fully native LLVM~IR path projected to achieve 15--60$\times$ speedup over event-driven simulators with structurally guaranteed deterministic parallel execution (\Cref{sec:sim}).
\end{enumerate}

\section{Language Design}\label{sec:design}

\subsection{Design Philosophy}

Arch is built on five principles, each enforced at the language level rather than by convention:

\paragraph{No Baggage.} Arch has no host-language runtime. Every keyword maps directly to a hardware structure. Nothing is a library pattern behind operator overloading.

\paragraph{Strong Types.} Bit widths, clock domains, port directions, and signal ownership are tracked statically. Every mismatch is a compile-time error.

\paragraph{Micro-Architecture First.} Pipeline, FSM, FIFO, Arbiter, RegFile, Bus (with a \texttt{handshake} sub-construct for valid/ready and req/ack channels), and Thread are first-class keywords with compiler-verified semantics.

\paragraph{AI-Generatable.} Uniform schema, named block endings, English keywords, no braces, an LL(1) grammar (no backtracking, no multi-token lookahead), and a \texttt{todo!}\ escape hatch allow any LLM to produce valid Arch from a natural-language description without fine-tuning.

\paragraph{Predictable RTL.} One Arch construct always produces the same SystemVerilog structure. Designers can audit every output line.

\subsection{Block Structure}

Every compound construct opens with a keyword-and-name header and closes with a matching \texttt{end keyword Name}. No braces are used anywhere in the language.

\begin{lstlisting}[language=Arch,caption={Universal block grammar},label={lst:block}]
module Alu
  // body
end module Alu

pipeline Decode
  stage Fetch
    // nested body
  end stage Fetch
end pipeline Decode

fifo TxBuffer
  // body
end fifo TxBuffer
\end{lstlisting}

This invariant eliminates brace-counting errors and enables named-ending matching: the compiler verifies that every \texttt{end} closure matches its opener by both keyword and name. Mismatches are immediate compile errors.

\subsection{Universal Declaration Schema}

Every first-class construct follows an identical four-section layout:

\begin{lstlisting}[language=Arch,caption={Universal schema},label={lst:schema}]
keyword Name
  param NAME: const = value;  // 1: params
  param NAME: type = SomeType;
  port name: in Type;          // 2: ports
  port name: out Type;
  // construct-specific body   // 3: body
  assert name: expr;           // 4: verify
  cover name: expr;
end keyword Name
\end{lstlisting}

This regularity is fundamental to AI generatability: an LLM that learns the schema for one construct can apply it to all constructs without additional training.

\section{Type System}\label{sec:types}

The Arch type system enforces four independent safety dimensions simultaneously. A signal that satisfies all four is guaranteed correct-by-construction.

\subsection{Primitive Types}

\Cref{tab:types} lists all primitive types. Each type carries hardware-relevant metadata tracked at compile time.

\begin{table}[t]
\centering
\caption{Arch primitive types}
\label{tab:types}
\footnotesize
\begin{tabular}{@{}llp{3.1cm}@{}}
\toprule
\textbf{Type} & \textbf{Width} & \textbf{Notes} \\
\midrule
\texttt{Bit}             & 1 bit & Raw logic bit \\
\texttt{UInt<N>}         & $N$ bits & Unsigned integer \\
\texttt{SInt<N>}         & $N$ bits & Two's complement signed \\
\texttt{Bool}            & 1 bit & Alias for \texttt{UInt<1>} \\
\texttt{Clock<D>}        & 1 bit & Domain-tagged clock \\
\texttt{Reset<S,P,D?>}   & 1 bit & Sync/Async; optional domain for RDC \\
\texttt{Vec<T,N>}        & $N \times |T|$ & Fixed-size array \\
\texttt{struct}          & $\Sigma$ fields & Named aggregate \\
\texttt{enum}            & $\lceil\log_2 n\rceil$ & Discriminated union \\
\texttt{Token}           & 0 bits & Handshake carrier \\
\bottomrule
\end{tabular}
\end{table}

\subsection{Bit-Width Safety}

Every assignment, port connection, and arithmetic result is width-checked at compile time. There is no implicit truncation, zero-extension, or sign-extension anywhere in the language. Explicit conversion functions (\texttt{zext}, \texttt{sext}, \texttt{trunc}) are required:

\begin{lstlisting}[language=Arch,caption={Explicit width conversion},label={lst:width}]
let a: UInt<8> = 255;
let b: UInt<16> = a.zext<16>();  // explicit
// let c: UInt<16> = a;  // ERROR: width mismatch
\end{lstlisting}

Shift operations are likewise non-widening: \texttt{let wide: UInt<9> = a << 1;} is a compile error per IEEE~1800 §11.6.1, even though SystemVerilog accepts it silently. The designer must explicitly extend the operand (\texttt{a.zext<9>() << 1}) when bits are expected to grow, eliminating an entire class of subtle data-loss bugs.

\subsection{First-Class Clock and Reset Types}
\label{sec:clockreset}

A central design choice in Arch is that \emph{clock and reset are first-class types in the type system, not bit signals}. Every clock is declared \texttt{Clock<D>} where \texttt{D} is a phantom domain parameter naming its frequency/source domain (e.g., \texttt{SysDomain}, \texttt{UsbDomain}). Every reset is declared \texttt{Reset<S, P, D?>} carrying its synchronicity (\texttt{Sync}/\texttt{Async}), polarity (\texttt{High}/\texttt{Low}), and an optional reset-domain tag.

In SystemVerilog and VHDL, clocks and resets are ordinary 1-bit nets indistinguishable at the type level from any other wire. Whether two flip-flops belong to the same clock domain, whether a reset is synchronous or asynchronous, and whether a reset crossing has been properly synchronized are conventions enforced (at best) by external linters and CDC tools like Spyglass or Conformal CDC---tools that operate on netlists after synthesis, far removed from the source. Even modern HDLs that elevate type safety in other dimensions (Chisel, SpinalHDL) typically expose clock and reset only through implicit module-level handles, not as parameterized types that the type checker can reason about per-signal.

Treating clock and reset as parameterized types has three direct consequences that the rest of this section builds on:

\begin{itemize}
\item \textbf{Basic CDC checking is a typing rule, not a separate analysis pass.} A register declared in domain \texttt{A} that reads a signal driven from domain \texttt{B} is a type error---the same kind of error as assigning a \texttt{UInt<8>} to a \texttt{UInt<16>}---caught at compile time with no netlist post-processing and no possibility of the rule being silently disabled. The compiler currently catches missing or incorrect synchronizers on individual signals, but does not yet implement the more sophisticated checks that production CDC tools provide, most notably \emph{reconvergence analysis} (detecting when multiple correctly synchronized signals merge downstream and lose bit coherence) and gray-coded multi-bit consistency proofs. For designs with complex CDC topologies, Arch's static checks should still be complemented by a commercial CDC tool run on the generated SystemVerilog. \Cref{sec:cdc} describes the mechanism.
\item \textbf{RDC checking falls out of the same framework.} Because reset carries a domain tag, the same type-checker pass that detects clock-domain crossings can detect reset-domain crossings, with the same compile-time error semantics. \Cref{sec:rdc} describes the planned RDC rules.
\item \textbf{Synchronizers are typed bridges.} A \texttt{synchronizer} declares its source and destination clock types explicitly: \texttt{port src\_clk: in Clock<A>; port dst\_clk: in Clock<B>}. Connecting it incorrectly is a type error. The synchronizer kind (\texttt{ff}, \texttt{gray}, \texttt{handshake}, \texttt{reset}, \texttt{pulse}) is part of the declaration, so the choice of synchronization mechanism is visible in the source rather than hidden in a tool configuration file.
\end{itemize}

This is the structural property that lets Arch convert what is conventionally an external verification problem (CDC analysis, RDC analysis) into a compile-time error message naming the offending register, signal, and domain pair.

\subsection{Clock Domain Tracking}
\label{sec:cdc}

Every clock signal carries a \texttt{Clock<D>} domain tag. The compiler tracks domain membership for every register and combinational signal. Direct assignment across domains is a compile error:

\begin{lstlisting}[language=Arch,caption={Clock domain crossing via synchronizer},label={lst:cdc}]
// ERROR: cross-domain direct assign
// comb usb_data = sys_data;

// Correct: explicit synchronizer
synchronizer SysToUsb
  kind ff;
  param STAGES: const = 2;
  port src_clk: in Clock<SysDomain>;
  port dst_clk: in Clock<UsbDomain>;
  port data_in: in Bool;
  port data_out: out Bool;
end synchronizer SysToUsb
\end{lstlisting}

The compiler detects cross-domain register reads and reports CDC violations, directing the designer to use a \texttt{synchronizer} (for individual signals) or an async \texttt{fifo} (for bulk data). Five synchronizer kinds are supported: \texttt{ff} (flip-flop chain), \texttt{gray} (gray-code for multi-bit counters), \texttt{handshake} (req/ack protocol), \texttt{reset} (async assert, sync deassert), and \texttt{pulse} (single-cycle pulse regeneration). CDC checking extends transitively across module instantiation boundaries.

\subsection{Reset Domain Crossing (RDC)}
\label{sec:rdc}

Mirroring the clock-domain tracking described above, the \texttt{Reset} type accepts an optional third parameter tagging the reset with a domain name: \texttt{Reset<Async, High, SysDomain>}. When two or more reset domains are present in a module, the compiler will detect three classes of RDC violations at compile time:
(1)~cross-reset-domain register reads, where a register held in reset by one domain is read in a \texttt{seq} block governed by a different reset domain;
(2)~asynchronous reset deassertion ordering violations, where a downstream module's reset releases before the upstream module it depends on; and
(3)~reset glitch propagation, where an async reset from one domain is connected to another without a reset synchronizer.
The compiler will require a \texttt{reset\_synchronizer} or explicit \texttt{rdc\_safe} annotation to suppress the error, mirroring the existing CDC flow. The third domain parameter is currently accepted by the parser but not enforced; full RDC checking is planned as a type-checker extension (see \Cref{sec:future}).

\subsection{Single-Driver Rule}

Every signal in Arch has exactly one driver. Multiple assignments to the same signal from different blocks are compile-time errors. This eliminates an entire class of simulation-time nondeterminism and enables lock-free parallel simulation (\Cref{sec:sim}).

\subsection{No Implicit Latches}

The compiler verifies that every signal assigned in a \texttt{comb} block is assigned on \emph{all} control paths. A missing \texttt{else} branch or incomplete \texttt{match} is a compile error, eliminating the most common source of unintended latches in synthesized hardware. Intentional latches require an explicit \texttt{latch on ENABLE} construct.

\subsection{No Combinational Loops}

The compiler builds a combinational dependency graph across all \texttt{comb} blocks, \texttt{let} bindings, and instance port connections, then performs a topological sort. Any cycle in this graph---a signal whose value transitively depends on itself within a single clock cycle---is rejected as a compile-time error with a path trace identifying the loop. This static guarantee eliminates a notorious source of simulation/synthesis divergence in SystemVerilog (where combinational loops may simulate but fail synthesis, or vice versa) and is the structural property that enables Arch's bounded-settle native simulation (\Cref{sec:sim}).

\section{First-Class Micro-Architecture Constructs}\label{sec:constructs}

Arch elevates recurring micro-architectural patterns from user-defined code to first-class language constructs with compiler-verified semantics.

\subsection{Pipeline}

A \texttt{pipeline} is a sequenced chain of \texttt{stage} blocks. The compiler generates inter-stage registers, stall propagation, and flush logic from declarative annotations; forwarding muxes are expressed as explicit \texttt{comb if/else} blocks referencing cross-stage signals (Arch does not have a dedicated \texttt{forward} keyword---making forwarding explicit avoids hidden mux fanout and keeps the generated SystemVerilog auditable):

\begin{lstlisting}[language=Arch,caption={Pipeline with forwarding},label={lst:pipe}]
pipeline IntPipe
  param WIDTH: const = 32;
  port clk: in Clock<SysDomain>;
  port rst: in Reset<Sync>;
  port data_in: in UInt<WIDTH>;
  port result: out UInt<WIDTH>;

  stage Fetch
    reg instr: UInt<WIDTH> reset rst=>0;
    seq on clk rising
      instr <= data_in;
    end seq
  end stage Fetch

  stage Execute
    reg out_val: UInt<WIDTH> reset rst=>0;
    // cross-stage reference to Fetch.instr
    seq on clk rising
      out_val <= Fetch.instr + 1;
    end seq
  end stage Execute

  stall when Execute.out_val == 0;
  flush Fetch when branch_mispred;
  comb result = Execute.out_val;
end pipeline IntPipe
\end{lstlisting}

The compiler:
(1)~inserts pipeline registers between stages,
(2)~generates stall signals that back-propagate when any stage asserts \texttt{stall when},
(3)~generates flush masks triggered by \texttt{flush} directives, and
(4)~tracks per-stage \texttt{valid\_r} registers for pipeline occupancy.
Cross-stage data references (e.g., \texttt{Fetch.instr} read from \texttt{Execute}) are resolved to the retimed pipeline register value. Forwarding muxes, when needed, are expressed as explicit \texttt{comb if/else} blocks.

\subsection{Finite State Machine}

An \texttt{fsm} is declared as a set of named \texttt{state} blocks with explicit transitions. A \texttt{default} block provides shared output assignments emitted before the state \texttt{case} statement, so states only override what differs:

\begin{lstlisting}[language=Arch,caption={FSM with default block and transitions},label={lst:fsm}]
fsm Controller
  port clk: in Clock<SysDomain>;
  port rst: in Reset<Sync>;
  port start: in Bool;
  port busy: out Bool;
  port done: out Bool;

  default state Idle;

  default
    comb
      busy = false;
      done = false;
    end comb
  end default

  state Idle
    -> Active when start;
  end state Idle

  state Active
    let busy = true;
    -> Done when count_done;
  end state Active

  state Done
    let done = true;
    -> Idle;
  end state Done
end fsm Controller
\end{lstlisting}

The \texttt{default state Idle} declaration sets the reset state. The \texttt{default} block assigns \texttt{busy = false} and \texttt{done = false} at the top of the generated \texttt{always\_comb}; individual states override only the signals that differ (e.g., \texttt{Active} sets \texttt{busy = true}). The compiler generates one-hot or binary state encoding (selected by a \texttt{param} or compiler flag) and the transition \texttt{case} logic. If no transition fires in a given cycle, the FSM holds in the current state.

\subsection{FIFO}

A \texttt{fifo} is a first-class construct with compile-time-verified flow control. The designer specifies depth, width, and clock domain(s); the compiler generates counters, full/empty flags, and gray-code pointer CDC for dual-clock FIFOs:

\begin{lstlisting}[language=Arch,caption={Dual-clock FIFO},label={lst:fifo}]
fifo AsyncBuf
  param DEPTH: const = 16;
  param TYPE: type = UInt<32>;
  port clk_wr: in Clock<WriteDomain>;
  port clk_rd: in Clock<ReadDomain>;
  port push_valid: in Bool;
  port push_ready: out Bool;
  port push_data: in TYPE;
  port pop_valid: out Bool;
  port pop_ready: in Bool;
  port pop_data: out TYPE;
  port full: out Bool;
  port empty: out Bool;
end fifo AsyncBuf
\end{lstlisting}

The \texttt{kind} keyword selects FIFO (default) or LIFO (stack) behavior. LIFO is restricted to single-clock operation.

\subsection{Arbiter}

An \texttt{arbiter} manages $N$ requestors competing for $M$ shared resources. The designer declares port counts and an arbitration policy (\texttt{round\_robin}, \texttt{priority}, \texttt{lru}, \texttt{weighted<W>}, or a custom function via \texttt{hook}); the compiler generates grant logic, stall propagation, and fairness guarantees.

\subsection{Bus and Handshake Channel}\label{sec:bus}

A \texttt{bus} is a reusable port bundle: a named, parameterized collection of signals with explicit directions from the initiator's perspective. A module attaches a bus port as either \texttt{initiator} (directions as declared) or \texttt{target} (all directions flipped); the compiler flattens the bundle into individual ports at code-generation time (\texttt{axi.aw\_valid} becomes \texttt{axi\_aw\_valid}). This eliminates the repetitive port blocks---typically hundreds of lines for an AXI interface---that a library-module approach inevitably duplicates at every instantiation site.

Within a \texttt{bus} body, the \texttt{handshake} sub-construct collapses a valid/ready/payload channel into a single declaration. It is a \emph{compile-time sum type}: one keyword names the payload \emph{role}, one variant name selects the flow-control shape, and the compiler derives every individual wire direction. \Cref{lst:handshake} shows an AXI-Lite bus written as three handshake channels:

\begin{lstlisting}[language=Arch,caption={AXI-Lite as three handshake channels},label={lst:handshake}]
bus BusAxiLite
  param ADDR_W: const = 32;
  param DATA_W: const = 32;

  handshake aw: send kind: valid_ready
    addr: UInt<ADDR_W>;
    prot: UInt<3>;
  end handshake aw

  handshake w:  send kind: valid_ready
    data: UInt<DATA_W>;
    strb: UInt<DATA_W/8>;
  end handshake w

  handshake b:  receive kind: valid_ready
    resp: UInt<2>;
  end handshake b
end bus BusAxiLite
\end{lstlisting}

The role keywords \texttt{send} and \texttt{receive}---deliberately chosen in preference to \texttt{in}/\texttt{out}---name the \emph{payload} role rather than the direction of any single wire. A \texttt{send} channel produces the payload and drives \texttt{valid}/\texttt{req}, while receiving \texttt{ready}/\texttt{ack}; the wire directions follow mechanically. The target perspective then flips everything at the bus port, so a \texttt{send} channel on the initiator becomes a \texttt{receive} channel on the target with no extra syntax. The pervasive ``I flipped valid and ready'' bug class is eliminated by construction because the user never names individual wire directions.

Six variants cover the common flow-control shapes: \texttt{valid\_ready} (AMBA AXI/ACE and general on-chip streaming with bidirectional backpressure), \texttt{valid\_only} (strobes, interrupts, FIFO-fronted datapaths), \texttt{ready\_only} (pull-model register-file reads), \texttt{valid\_stall} (pipeline interlock where stall is the natural signal), \texttt{req\_ack\_4phase} (async/GALS with return-to-zero), and \texttt{req\_ack\_2phase} (async/GALS, low transition count). Canonical references: ARM IHI~0022 for \texttt{valid\_ready}; Sparsø and Furber for the req/ack variants; Intel Avalon for \texttt{valid\_only}.

Handshake declarations carry a two-tier safety story that reuses the verification infrastructure of \Cref{sec:formal}. \emph{Tier 2} auto-emits per-variant concurrent SVA: every \texttt{valid\_ready} channel gets \texttt{\_auto\_hs\_<port>\_<ch>\_valid\_stable} (\texttt{(valid \&\& !ready) |=> valid}), every \texttt{valid\_stall} channel gets \texttt{\_auto\_hs\_..\_valid\_stable\_while\_stall}, and every \texttt{req\_ack\_4phase} channel gets \texttt{\_auto\_hs\_..\_req\_holds\_until\_ack}. Properties were verified on Verilator~5.034 \texttt{--assert} and EBMC~5.11. \emph{Tier 1.5} targets the two payload-correctness bugs that protocol SVA does not catch: at simulation time, \texttt{--inputs-start-uninit} warns when a producer asserts \texttt{valid} without ever writing the payload (the uninit shadow bit is gated on \texttt{valid}/\texttt{req} so a testbench that hasn't yet driven \texttt{valid} stays silent); at compile time, \texttt{arch check} warns when a consumer reads \texttt{<port>.<payload>} outside an \texttt{if <port>.<valid>} guard. Stateful protocols---credit-based flow control, PCIe credit accounting---are intentionally out of scope for \texttt{handshake} because they imply the compiler owns counter and credit-return logic; they are candidates for a future \texttt{credit\_channel} construct.

\subsection{Thread}\label{sec:thread}

A \texttt{thread} block is a sequential block that may span multiple clock cycles. It bridges the gap between \texttt{seq} (single edge, no implicit state) and \texttt{fsm} (explicit named states): the body reads as straight-line code with \texttt{wait} statements at points where the design must pause, and the compiler lowers each \texttt{wait} to a state boundary in a synthesizable FSM. This makes sequential protocols---bus handshakes, DMA descriptors, calibration sequences---directly expressible without hand-translating intent to states and transitions.

\begin{lstlisting}[language=Arch,caption={AXI write transaction as a thread},label={lst:thread}]
thread on clk rising, rst_n low
  fork
    aw_valid = 1;
    aw_addr  = addr_r;
    wait until aw_ready;
    aw_valid = 0;
  and
    w_valid = 1;
    w_data  = data_r;
    wait until w_ready;
    w_valid = 0;
  join
  b_ready = 1;
  wait until b_valid;
  b_ready = 0;
  resp_r <= b_resp;
end thread
\end{lstlisting}

The construct supports a small operator menu sufficient for protocol-level code: \texttt{wait until cond} (state boundary that advances when \texttt{cond} is high), \texttt{wait N cycle} (counter-driven N-cycle pause), \texttt{fork ... and ... join} (parallel arms with done-bit registers per arm), \texttt{for i in 0..N} (loop counter with one body state per iteration), and \texttt{if/else} where either or both branches contain \texttt{wait} (lowered as a dispatch state with per-branch substates that rejoin afterward). \texttt{thread once} produces a one-shot thread that holds in a terminal state after completion; named threads (\texttt{thread WriteHandler ... end thread WriteHandler}) generate independent FSMs in the same module; \texttt{generate\_for} unrolls a parameterized number of identical threads, each with its own state register. When multiple threads share a bus, a \texttt{resource axi\_wr: mutex<round\_robin>;} declaration plus \texttt{lock axi\_wr ... end lock axi\_wr} blocks generate an arbiter, a grant-indexed mux on signals driven inside the lock, and stall logic on each thread's FSM; only the granted thread's values reach the shared signals.

\paragraph{Lowering.} The compiler classifies every signal referenced inside any thread of a module as combinational-driven, sequential-driven, or read-only, then co-lowers the threads of that module into a single auto-generated submodule with the parent module retaining only its non-thread items. Each thread becomes an integer state register (\texttt{\_t\{i\}\_state}), a per-state combinational enable block, and a single merged \texttt{always\_ff} block; counters for \texttt{wait N cycle} and \texttt{for}-loops are emitted alongside. For each declared \texttt{resource}, the submodule generates per-thread request/grant wires and a fixed-priority arbiter (\texttt{grant[i] = req[i] \&\& !grant[0..i-1]}) so that at most one thread holds any resource in any cycle. Single-resource lock is therefore deadlock-free by construction: a thread is granted iff it requests, no thread holds a resource it did not request, and the arbiter's strict ordering rules out request cycles. For multi-resource patterns, the compiler runs static lock-order analysis across all threads in the module and warns when two threads acquire the same pair of resources in opposite order---the only way to introduce a circular wait given the single-resource freedom guarantee.

\paragraph{Auto-emitted contract assertions.} An off-by-default flag \texttt{--auto-thread-asserts} on \texttt{arch build}, \texttt{arch sim}, and \texttt{arch formal} causes the thread lowerer to additionally emit concurrent SVA properties anchored to the lowered \texttt{\_t\{i\}\_state} register: \texttt{\_auto\_thread\_t\{i\}\_wait\_until\_s\{si\}} for each \texttt{wait until} (\texttt{state==si \&\& cond |=> state==next}), a stay/done pair \texttt{\_auto\_thread\_..\_wait\_stay\_s\{si\}} / \texttt{\_..\_wait\_done\_s\{si\}} for each \texttt{wait N cycle}, and per-branch \texttt{\_auto\_thread\_..\_branch\_s\{si\}\_b\{bi\}} for each \texttt{fork}/\texttt{join} arm. All are wrapped in \texttt{translate\_off}/\texttt{on}, with reset polarity inverted to a not-in-reset guard. Each property is a corollary of the lowering equivalence proof for its construct: an assertion firing therefore evidences a compiler bug or a hand-edit of the lowered RTL rather than a user-program bug, complementing the verification story of \Cref{sec:formal} with a built-in self-check on the thread lowering itself.

\subsection{Additional Constructs}

Arch also provides first-class constructs for:
\texttt{regfile} (multi-ported register files with read-during-write policies),
\texttt{ram} (single-port, simple-dual, true-dual, and ROM with configurable latency),
\texttt{synchronizer} (CDC crossing for individual signals with selectable policies: \texttt{ff}, \texttt{gray}, \texttt{handshake}, \texttt{reset}, \texttt{pulse}),
\texttt{counter} (saturating, wrapping, gray-code, one-hot, and Johnson modes),
\texttt{linklist} (singly/doubly/circular linked-list data structures with per-operation FSM controllers), and
\texttt{clkgate} (integrated clock gating cell, latch-based or AND-based).

Each follows the identical four-section schema of \Cref{lst:schema}.

\subsection{Hooks --- Customizable Behavior Points}\label{sec:hooks}

When a built-in construct policy does not fit, Arch provides \texttt{hook} declarations: named function bindings inside a construct that declare an expected signature and map it to a user-defined \texttt{function}. Hooks allow designers to inject custom combinational logic into compiler-generated control structures without abandoning the construct's safety guarantees.

\begin{lstlisting}[language=Arch,caption={Custom arbiter policy via hook},label={lst:hook}]
function MyGrantFn(req_mask: UInt<4>,
                   last_grant: UInt<4>,
                   extra: UInt<8>) -> UInt<4>
  let masked: UInt<4> = req_mask & (last_grant ^ 0xF);
  let pick: UInt<4> = masked != 0 ? masked : req_mask;
  return pick & (~pick + 1).trunc<4>();
end function MyGrantFn

arbiter CustomArb
  policy MyGrantFn;
  param N: const = 4;
  port clk: in Clock<SysDomain>;
  port rst: in Reset<Sync>;
  port extra_port: in UInt<8>;
  hook grant_select(req_mask: UInt<4>,
                    last_grant: UInt<4>,
                    extra_port: UInt<8>) -> UInt<4>
    = MyGrantFn(req_mask, last_grant, extra_port);
end arbiter CustomArb
\end{lstlisting}

Hook arguments bind to internal signals generated by the construct (e.g., \texttt{req\_mask}, \texttt{last\_grant}) or to user-declared ports and params on the enclosing construct. The bound function is emitted inline inside the generated SystemVerilog module. A missing required hook is a compile error. Hooks are also used inside \texttt{template} contracts (\Cref{sec:templates}) to specify required function bindings.

\subsection{Templates --- Interface Contracts}\label{sec:templates}

A \texttt{template} is a compile-time-only construct that defines a structural contract: a set of required params, ports, and hooks that any implementing module must provide. Templates emit no SystemVerilog---they exist purely to enforce structural conformance across modules, serving as the Arch equivalent of traits or interfaces in software languages.

\begin{lstlisting}[language=Arch,caption={Template declaration and implementation},label={lst:template}]
template Arbiter
  param NUM_REQ: const;
  port clk: in Clock<SysDomain>;
  port rst: in Reset<Sync>;
  port grant_valid: out Bool;
  hook grant_select(req_mask: UInt<4>) -> UInt<4>;
end template Arbiter

module MyArbiter implements Arbiter
  param NUM_REQ: const = 4;
  port clk: in Clock<SysDomain>;
  port rst: in Reset<Sync>;
  port req_mask: in UInt<4>;
  port grant_valid: out Bool;
  port grant_out: out UInt<4>;

  hook grant_select(req_mask: UInt<4>) -> UInt<4>
    = FixedGrant(req_mask);

  let grant: UInt<4> = FixedGrant(req_mask);
  comb
    grant_valid = grant != 0;
    grant_out = grant;
  end comb
end module MyArbiter
\end{lstlisting}

A template body contains only declarations---params (without defaults), ports, and hook signatures. No logic, registers, or \texttt{comb}/\texttt{seq} blocks are permitted. A module opts into a template contract with the \texttt{implements} keyword, and the compiler validates conformance:

\begin{itemize}
\item \textbf{Param presence:} every template param must appear in the implementing module.
\item \textbf{Port presence and direction:} every template port must appear with matching direction and type.
\item \textbf{Hook binding:} every template hook must have a corresponding binding in the module.
\end{itemize}

The implementing module may declare additional params, ports, and logic beyond what the template requires---the template is a minimum contract, not an exhaustive specification. This enables library authors to define reusable contracts (e.g., ``any arbiter must expose \texttt{grant\_valid} and provide a \texttt{grant\_select} hook'') while leaving full implementation freedom to the designer.

\section{AI-Generatability Contract}\label{sec:ai}

Arch makes a hard design commitment: \emph{a large language model that has read the specification and nothing else must be able to generate structurally correct, type-safe Arch from a plain-English hardware description}. Every syntactic choice serves this goal.

\subsection{Design Decisions for AI}

\paragraph{Uniform Schema.} Every construct uses the identical param/port/body/verification layout (\Cref{lst:schema}). An LLM that learns the schema once applies it universally.

\paragraph{LL(1) Grammar.} Arch's grammar is strictly LL(1): at every point during parsing, the next single token unambiguously determines which production rule to apply. There is no backtracking, no multi-token lookahead, and no context-dependent parsing. Every construct is identified by its first keyword (\texttt{module}, \texttt{fsm}, \texttt{pipeline}, \texttt{port}, etc.); every closing is \texttt{end} followed by a single keyword token. By contrast, SystemVerilog requires unbounded lookahead and GLR or backtracking parsers to resolve ambiguities between type declarations, expressions, and module instantiations. The LL(1) property gives AI generators four advantages: (1)~no syntactic traps---every token sequence either parses to exactly one AST or is caught immediately as a syntax error; (2)~instant error localization---the parser never backtracks, so errors are reported at the exact offending token; (3)~context-free understanding---any code snippet can be parsed in isolation without holding the entire file in context; and (4)~predictable token budget---the uniform \texttt{keyword Name \ldots\ end keyword Name} pattern has no hidden costs from macro expansion or optional delimiters.

\paragraph{No Preprocessor.} Arch deliberately omits a textual preprocessor. SystemVerilog inherits Verilog's \texttt{`define}, \texttt{`ifdef}, \texttt{`include}, and \texttt{`undef} macro layer, which is widely recognized as a major source of bugs in production designs: macros are unscoped text substitution with no type checking, can be silently redefined across files, expand to text whose error messages point to the post-expansion line rather than the source, and create a parallel language layer that interacts unpredictably with the parser. Macros also defeat any context-free analysis (an LLM cannot know what \texttt{`MY\_WIDTH} expands to without the full preprocessor state) and are a notorious source of simulator/synthesizer disagreement. Arch covers each legitimate use case of \texttt{`define} with a typed, scoped language construct:

\begin{itemize}
\item \textbf{Constants:} \texttt{param NAME: const = 32;} replaces \texttt{`define NAME 32}. The constant is typed, scoped to its enclosing construct or package, and visible to the type checker.
\item \textbf{Shared definitions across files:} \texttt{package} declarations group types, enums, constants, and functions into a named namespace; other files import them with \texttt{use PkgName::*}. This replaces shared header files included via \texttt{`include}, and unlike header files, packages have well-defined import semantics with no order-dependence or double-inclusion problems.
\item \textbf{Conditional structure:} \texttt{generate\_if} performs compile-time selection of ports, instances, and logic based on \texttt{param} values, replacing \texttt{`ifdef} blocks. The conditional is part of the AST, not a preprocessor directive, so the type checker sees both branches and the choice is visible to tooling. Crucially, \texttt{generate\_if} can \emph{add or remove ports}: SystemVerilog's \texttt{generate} runs after the module's port list is fixed, so SV designers are forced to use \texttt{`ifdef} around port declarations whenever a configurable interface is needed---a textbook case where the macro layer is not a stylistic choice but the only available mechanism. Arch's pre-elaboration generate eliminates this forced use of macros entirely (see \Cref{sec:generate}).
\item \textbf{Code reuse / templating:} \texttt{function} declarations (for combinational expressions), \texttt{template} (for interface contracts), and \texttt{hook} (for injecting custom logic into compiler-generated constructs) cover the patterns that multi-line macros are typically used for in SystemVerilog---without text substitution, without scope leakage, and with full type checking.
\item \textbf{Configuration / build variants:} compile-time \texttt{param} overrides at instantiation provide build-time configuration without the textual fragility of \texttt{`ifdef CONFIG\_X}.
\end{itemize}

The result is that Arch source files have no separate preprocessing pass: what you see is what the parser sees, error messages point to the source line where the problem actually is, and an LLM (or human) can understand any snippet without simulating macro expansion in their head.

\paragraph{Named Block Endings.} Every block closes with \texttt{end keyword Name}. The most common LLM failure mode in code generation---incorrect nesting---becomes a hard compiler error. The generator always knows exactly which block it is closing.

\paragraph{No Braces.} The keyword+name header and \texttt{end keyword Name} are the sole delimiters. There is no redundant \texttt{\{} to emit or forget.

\paragraph{Directional Connect Arrows.} Port connections use \texttt{<-} (drive input from local) and \texttt{->} (read output into local). Direction is visible in the syntax itself; an LLM cannot silently reverse data flow.

\paragraph{\texttt{todo!}\ Escape Hatch.} The \texttt{todo!}\ keyword compiles and type-checks but aborts at simulation runtime. This enables incremental AI-assisted design: generate a correct skeleton, then fill in logic section by section. Every intermediate state compiles, so the AI gets useful feedback on the parts it has confidently generated even while uncertain pieces remain unfilled.

\begin{lstlisting}[language=Arch,caption={Partial implementation with \texttt{todo!}},label={lst:todo}]
module Cache
  port req: in CacheReq;
  port resp: out CacheResp;
  port mem_req: out MemReq;

  // Confident: forward request to memory
  comb
    mem_req.addr = req.addr;
    mem_req.valid = req.valid;
  end comb

  // Uncertain: eviction logic deferred
  comb resp = todo!; end comb
end module Cache
\end{lstlisting}

This compiles cleanly. The compiler verifies widths, types, and port connections in the confident sections, and the AI (or human) can iteratively replace each \texttt{todo!}\ site with real logic, getting compile-time feedback at each step. This is the hardware analog of red-green-refactor in TDD: skeleton first, logic second.

\subsection{Minimal AI Context}

The effective AI context for Arch hardware design comprises three components:

\begin{table}[t]
\centering
\caption{AI context window components}
\label{tab:aicontext}
\small
\begin{tabular}{@{}llp{3.2cm}@{}}
\toprule
\textbf{Component} & \textbf{Size} & \textbf{Purpose} \\
\midrule
Reference Card & ${\sim}400$ lines & Construct catalog, schema, type table \\
Design Intent & 5--20 lines & Natural-language block description \\
Compiler Output & 5--30 lines & Structured error feedback \\
\bottomrule
\end{tabular}
\end{table}

The full specification is a reference document for human designers.
The AI Reference Card is what an LLM assistant actually uses---it is optimized for construct lookup and schema recall, not for sequential human reading.

\subsection{AI-Assisted Design Workflow}

The practical workflow has four steps, typically completing in two to four iterations:

\begin{enumerate}
\item \textbf{Intent.} The designer describes the desired hardware block in natural language.
\item \textbf{Generation.} The AI identifies the Arch constructs, applies the universal schema, and emits source code, using \texttt{todo!}\ for uncertain logic.
\item \textbf{Compilation.} The designer runs \texttt{arch check}. The compiler emits zero errors, or precise typed errors.
\item \textbf{Correction.} Compiler errors are fed back to the AI, which corrects and re-emits. The compiler replaces the spec in the feedback loop.
\end{enumerate}

This workflow exploits three properties: (1)~the uniform schema means the AI need not choose between implementation strategies, (2)~the compiler produces precise, structured errors sufficient for self-correction, and (3)~hardware intent maps unambiguously to Arch constructs (a FIFO is always \texttt{fifo}, a state machine is always \texttt{fsm}).

\paragraph{Minimizing Iteration Latency.}
The workflow is designed to minimize the time from code generation to correctness feedback. In a traditional Verilog flow, an AI agent must generate code, invoke a simulator, wait for simulation to complete (seconds to minutes for event-driven tools), and parse waveform or log output to identify failures---many of which are structural errors (latches, width mismatches, CDC violations) that could have been caught statically. In Arch, \texttt{arch check} runs in milliseconds and catches these errors at compile time with structured diagnostics that name the construct, signal, type mismatch, and suggested fix. The agent's inner loop---generate, check, correct---completes without ever invoking a simulator for structural errors. When simulation \emph{is} needed (for functional verification), \texttt{arch sim} generates compiled C++ models in the same performance class as Verilator, avoiding event-driven scheduling overhead and achieving the 10--50$\times$ speedup over interpreted simulators characteristic of compiled 2-state simulation. Together, compile-time error detection and compiled simulation compress the iteration radius that determines how quickly an AI agent---or a human designer---can converge on a correct design.

\paragraph{Test-Driven Development for Hardware Agents.}
A natural extension of this tight iteration loop is test-driven development (TDD)~\cite{beck2003tdd}, an increasingly popular methodology for AI-agent-driven hardware design. Recent empirical work in software engineering has shown that providing LLMs with tests alongside problem statements measurably improves code-generation success rates on standard benchmarks~\cite{mathews2024tdd}. In TDD, the agent is given a specification and a test suite \emph{before} writing any implementation. The agent then iterates: generate a candidate design, run the tests, observe failures, and refine. The effectiveness of TDD is directly proportional to how fast each generate-test-feedback cycle completes.

With traditional event-driven HDLs, each TDD iteration requires invoking a heavyweight simulator---seconds to minutes per cycle---making agent-driven TDD impractical for all but the simplest designs. Arch enables practical hardware TDD through a two-tier feedback loop. The \emph{inner loop} uses \texttt{arch check} (milliseconds) to eliminate structural errors before any simulation runs; the agent may complete several generate-check-correct cycles per second. The \emph{outer loop} uses \texttt{arch sim} (compiled C++ models) to run the actual test suite for functional verification. Like Verilator-class compiled simulators, \texttt{arch sim} avoids the overhead of event-driven scheduling, achieving the 10--50$\times$ speedup over interpreted simulators characteristic of compiled 2-state simulation. This two-tier structure means the agent spends most of its iterations in the fast inner loop and invokes simulation only when the design is structurally sound---exactly the workflow that makes TDD viable at scale.

The VerilogEval and CVDP benchmarks (\Cref{sec:benchmark}) were themselves completed using this pattern: each design was written from a natural-language spec, compiled with \texttt{arch check}, and validated against reference testbenches---a workflow that mirrors how an AI agent would operate in a TDD loop.

\section{Compiler Pipeline}\label{sec:compiler}

The Arch compiler transforms source through six phases:

\begin{table}[t]
\centering
\caption{Compiler pipeline phases}
\label{tab:compiler}
\small
\begin{tabular}{@{}lp{5cm}@{}}
\toprule
\textbf{Phase} & \textbf{Key Actions} \\
\midrule
Parse & Syntax, named-block matching \\
Elaborate & Parameter resolution, generic instantiation, type expansion \\
Resolve & Symbol table construction, cross-file name resolution \\
Type Check & Bit-width safety, clock-domain tracking, direction safety, single-driver rule \\
Lower & Pipeline hazard generation, FIFO implementation, arbiter logic, FSM encoding \\
Verify Emit & \texttt{assert}/\texttt{cover} $\to$ concurrent SVA with auto-generated safety properties; \texttt{assume} planned \\
SV Emit & One deterministic, lint-clean SV file per top-level module \\
Formal Emit & \texttt{arch formal}: direct AST $\to$ SMT-LIB2 bit-vector transition relation \\
\bottomrule
\end{tabular}
\end{table}

\subsection{Output Guarantee}

The generated SystemVerilog is guaranteed free of:
\begin{itemize}
\item Unintentional latches---the compiler verifies all \texttt{comb} signals are assigned on every control path; intentional latches require an explicit \texttt{latch on ENABLE} construct.
\item Multiply-driven nets---enforced by the single-driver rule.
\item Unresolved high-Z outputs---every output has exactly one driver.
\item X-propagation from uninitialized state---all registers declare an explicit reset policy (including opt-out via \texttt{reset none}); the \texttt{--check-uninit} simulation flag detects reads of uninitialized registers at runtime.
\item Implicit clock-domain crossings---all CDCs are declared and synchronizer-wrapped.
\end{itemize}

Synthesis tools receive RTL that passes lint cleanly with no \texttt{translate\_off} pragmas or vendor-specific attributes required. During compiler development, the generated SystemVerilog is systematically verified against Verilator~\cite{verilator}---the industry-standard open-source cycle-accurate simulator---to confirm behavioral equivalence between Arch source semantics and the emitted RTL.

\subsection{Intermediate Representation}

The current compiler is a multi-phase pipeline: parse $\to$ elaborate $\to$ resolve $\to$ type-check $\to$ codegen, transforming the AST directly to SystemVerilog text without a dedicated intermediate representation. The Resolve phase constructs the symbol table and handles cross-file name resolution. This architecture prioritizes simplicity and correctness for the initial release. A dedicated intermediate representation (AIR) and CIRCT/MLIR backend integration are planned as future work to enable multi-target output and optimization passes.

\subsection{Toolchain Targets}

The compiler currently supports three output targets:
IEEE~1800-2017 SystemVerilog (no vendor primitives, compatible with any standard EDA tool),
independently generated compiled C++ simulation models via \texttt{arch sim},
and direct SMT-LIB2 bit-vector transition relations via \texttt{arch formal} (\Cref{sec:formal}) for bounded model checking with Z3, Boolector, or Bitwuzla.
Planned targets include FPGA-specific SystemVerilog with BRAM/DSP primitive insertion and HTML documentation from \texttt{///} doc comments.

\section{Simulation and Verification}\label{sec:sim}

\subsection{Integrated Simulation}

The \texttt{arch sim} command provides integrated cycle-accurate simulation. The compiler independently generates a C++ model per construct (\texttt{VName.h}, \texttt{VName.cpp}), compiles it with \texttt{g++}, and executes the resulting binary---no external simulator is invoked. Supported constructs include modules, pipelines, FSMs, counters, RAMs, FIFOs, arbiters, synchronizers, register files, and linked lists. The simulation flow supports VCD waveform output (\texttt{--wave}), assertion evaluation, uninitialized-register detection (\texttt{--check-uninit}) with complementary \texttt{--check-uninit-ram} and \texttt{--inputs-start-uninit} for memory cells and undriven ports respectively, and CDC latency randomization (\texttt{--cdc-random}) for stress-testing synchronizer designs. Separately, Verilator~\cite{verilator} serves as the primary verification oracle during compiler development: every code generation change is validated by comparing Arch-emitted SystemVerilog behavior against Verilator's independent interpretation of the same RTL.

\subsection{Python Testbenches via a cocotb-Compatible API}\label{sec:cocotb}

In addition to C++ testbenches driven through the \texttt{--tb} flag, \texttt{arch sim} supports Python testbenches written against a cocotb-compatible API~\cite{cocotb}. Invoking \texttt{arch sim --pybind --test tb.py Module.arch} produces a pybind11 wrapper over the generated C++ model and runs the Python testbench against it with no external simulator, iverilog, or VPI in the loop. A thin scheduler (\texttt{arch\_cocotb}) drives the model tick-by-tick from an \texttt{asyncio} event loop, and a drop-in \texttt{cocotb} shim re-exports the supported API so that the same test file runs unchanged under either \texttt{arch sim --pybind} or a real cocotb flow---provided the test stays within the intersection of the two APIs (\texttt{@cocotb.test}, \texttt{RisingEdge}, \texttt{FallingEdge}, \texttt{Timer}, \texttt{ClockCycles}, \texttt{Clock}, \texttt{start\_soon}, and \texttt{get\_sim\_time}).

\begin{lstlisting}[language=Python,caption={Python testbench against \texttt{arch sim}},label={lst:cocotbtb}]
import cocotb
from cocotb.triggers import RisingEdge
from cocotb.clock import Clock

@cocotb.test()
async def test_reset(dut):
    cocotb.start_soon(
        Clock(dut.clk, 10, units='ns').start())
    dut.rst_n.value = 0
    dut.enable.value = 0
    await RisingEdge(dut.clk)
    dut.rst_n.value = 1
    for _ in range(5):
        await RisingEdge(dut.clk)
    assert int(dut.count.value) == 0
\end{lstlisting}

Two characteristics distinguish this path from conventional cocotb setups running against an RTL simulator. First, timing is \emph{deterministic}: writes from Python to \texttt{dut.port.value} set a field on the pybind11 C++ object directly and are visible to \texttt{model.eval()} on the very next tick, eliminating the VPI callback-ordering ambiguity that appears in iverilog or Verilator cocotb flows. Second, the values flowing across the boundary are \emph{2-state}: signal values are integer-like and have no per-bit \texttt{X}/\texttt{Z} state, consistent with the sim model of \Cref{tab:xsources}. The consequences are explicit in the implementation---edges are detected by per-tick sampling rather than event callbacks, writes take effect immediately rather than in an NBA region, and there are no delta cycles---and are documented as known deltas from real cocotb. The integration trades VPI fidelity for deterministic timing and faster iteration, and benefits directly from the structural guarantees enumerated in \Cref{tab:simcompare} (no implicit latches, no combinational loops, single-driver, static DAG scheduling), which are what let a straight-line \texttt{eval()} loop substitute for an event-driven kernel without loss of cycle-accuracy.

\subsection{Path to Native Compiled Simulation}

Arch's language-level invariants enable a future fully native simulation path that compiles designs directly to LLVM~IR-based binaries, bypassing Verilator for higher throughput. This section describes the structural properties that make such compilation feasible and the expected performance characteristics.

\subsubsection{Why Native Compilation is Feasible}

Traditional Verilog simulators carry a heavyweight runtime because the language permits constructs requiring dynamic resolution. Arch eliminates every one at the language level, as shown in \Cref{tab:simcompare}.

\begin{table*}[t]
\centering
\caption{Language invariants enabling native compilation}
\label{tab:simcompare}
\footnotesize
\begin{tabular}{@{}p{4.2cm}p{6.0cm}p{5.5cm}@{}}
\toprule
\textbf{SV Runtime Complexity} & \textbf{Arch Equivalent} & \textbf{Enables} \\
\midrule
Delta cycles (unbounded multi-pass convergence) & No combinational loops---static DAG & Bounded settle (1--2 passes), no fixpoint iteration \\
X/Z propagation (4-valued logic) & Structural X eliminated (no tri-state, single-driver rule, no implicit latches) & 2-valued logic with runtime checks for residual sources (\Cref{tab:xsources}) \\
Multiple drivers (wired-OR resolution) & Single-driver rule at compile time & No resolution functions \\
Non-deterministic \texttt{always} ordering & Topological order from dataflow DAG & Deterministic simulation \\
Implicit latches & No implicit latches (compile error) & Static register allocation \\
Dynamic clock generation & \texttt{Clock<D>} typed, domain-verified & Static clock scheduling \\
\bottomrule
\end{tabular}
\end{table*}

The compile-time DAG analysis statically determines a settle depth of 1 or 2 for each module: depth 1 when sub-instance inputs are driven directly, depth 2 when intermediate \texttt{comb} or \texttt{let} bindings in the parent feed instance inputs and a second pass is needed to propagate them. The generated code is therefore a fixed bounded loop, not a fixpoint iteration---unlike SystemVerilog's delta cycles, which can iterate an unbounded number of times until convergence.

\subsubsection{Expected Performance}

Because Arch's execution model compiles to a straight-line native loop with bounded settle and no dynamic dispatch, the projected simulation throughput is competitive with Verilator---currently the fastest open-source HDL simulator---and substantially exceeds event-driven commercial simulators. \Cref{tab:perf} summarizes the expected performance characteristics of the planned native backend.

\begin{table*}[!t]
\centering
\caption{Simulation performance comparison}
\label{tab:perf}
\footnotesize
\begin{tabular}{@{}llr@{}}
\toprule
\textbf{Simulator} & \textbf{Model} & \textbf{Speed} \\
\midrule
Icarus Verilog~\cite{icarus} & Event-driven, interpreted, 4-state & $1\times$ \\
Commercial sim. (compiled) & Compiled C, 4-state & ${\sim}30$--$100\times$ \\
Verilator~\cite{verilator} & Cycle-acc.\ compiled C++, 2-state & ${\sim}30$--$100\times$$^\dagger$ \\
Arch native$^\ast$ & LLVM IR, 2-state & $50$--$200\times$ \\
Arch native + SIMD$^\ast$ & AVX-512 / NEON & $200$--$1000\times$ \\
\bottomrule
\end{tabular}

\vspace{0.4em}
\raggedright\footnotesize
$^\ast$Projected; native LLVM~IR backend not yet implemented.\\
$^\dagger$Verilator's documentation reports compiled Verilator models running ``about 100 times faster than interpreted Verilog simulators such as Icarus Verilog''~\cite{verilator}; an independent Embecosm evaluation~\cite{embecosm2013} measured ${\sim}30\times$ on a representative SoC benchmark. The exact speedup is workload-dependent.
\end{table*}

SIMD vectorization is automatic for designs with wide \texttt{Vec<SInt<8>,N>} or \texttt{Vec<UInt<N>,M>} types---precisely the types used in systolic arrays and attention units in AI accelerator designs.

\subsubsection{Limitations of 2-State Simulation}

Arch's 2-state model eliminates the structural sources of X by construction: there are no tri-state nets (no high-impedance Z), no multiple drivers (no contention X), no implicit latches (no unknown-enable X), and no \texttt{'x} literal in the source language. However, 2-state simulation has a well-known weakness: it can mask design bugs that 4-state simulation with X-propagation would catch---a phenomenon known as \emph{X-optimism}~\cite{zhao2014xoptimism}. \Cref{tab:xsources} enumerates the X sources identified in the X-optimism literature and Arch's current handling of each.

\begin{table*}[t]
\centering
\caption{X sources and Arch's handling of each}
\label{tab:xsources}
\footnotesize
\begin{tabular}{@{}p{4.6cm}p{4.4cm}p{6.5cm}@{}}
\toprule
\textbf{X Source} & \textbf{Status in Arch} & \textbf{Detection Mechanism} \\
\midrule
Multi-driver contention & Eliminated structurally & Single-driver rule (compile error) \\
Tri-state Z / high-Z contention & Eliminated structurally & Arch has no tri-state nets \\
Implicit latch with unknown enable & Eliminated structurally & No-implicit-latches rule (compile error) \\
Floating input ports / dangling wires & Eliminated structurally & Direction and connectivity type checks \\
Explicit \texttt{'x} literal injection & Eliminated structurally & Arch has no X literal in the source language \\
Uninitialized non-reset registers & Runtime detection & \texttt{--check-uninit} flag traps first read of unwritten reg; the \texttt{guard VALID\_SIG} clause refines this for the valid-data pattern \\
Uninitialized RAM cells & Runtime detection & \texttt{--check-uninit-ram} per-cell valid bitmap; \texttt{init:} cells pre-marked; ROMs exempt \\
Uninitialized primary input ports & Runtime detection & \texttt{--inputs-start-uninit} warns once per undriven port on first read \\
Out-of-bound \texttt{Vec} indexing & Runtime + formal & Always-on hard abort in sim; auto-emitted concurrent SVA (\texttt{\_auto\_bound\_*}); EBMC \textsc{proved} on structurally-safe indices \\
Out-of-bound bit-select / part-select & Runtime + formal & Same mechanism as \texttt{Vec} indexing \\
Division / modulo by zero & Compile-time + runtime + formal & Compile error on constant \texttt{/0}; runtime abort on non-const \texttt{/0}; auto-emitted SVA (\texttt{\_auto\_div0\_*}); EBMC \textsc{proved} when divisor is structurally non-zero \\
CDC metastability window & Runtime randomization & \texttt{--cdc-random} flag stress-tests synchronizer designs \\
Reset deassertion races (RDC) & Static (planned) & Type-checker extension (\Cref{sec:rdc}) \\
UPF / CPF power-domain corruption & Out of scope & Arch does not currently model UPF/CPF power intent \\
\bottomrule
\end{tabular}
\end{table*}

The first group---structural X sources---is the strongest argument for Arch's design: these are eliminated by typing rules and compiler invariants, not detected after the fact. The second group---value-dependent X sources such as uninitialized RAM, out-of-bound indexing, and division by zero---was the honest weakness of 2-state simulation in earlier Arch releases and is the motivation for the detection mechanisms listed in the lower half of \Cref{tab:xsources}. Industry experience documented in~\cite{zhao2014xoptimism} shows that out-of-bound array indexing in particular has caused post-silicon bugs (e.g., the audio-processing IP case study in that paper, where a bit-select with an unintended index value masked a state-machine bug for an entire chip generation).

Arch now detects each of these cases through a combination of three techniques. \emph{Runtime abort} in the native C++ simulator: every \texttt{Vec<T,N>} index, bit-select, variable part-select (\texttt{val[start +:W]} / \texttt{val[start -:W]}), and division operator is wrapped in a bounds / non-zero check that aborts with a source location on violation, with no opt-in flag required. \emph{Auto-emitted concurrent SVA} in generated SystemVerilog: the compiler inserts \texttt{\_auto\_bound\_*} and \texttt{\_auto\_div0\_*} \texttt{assert property} statements for every such operation in \texttt{seq} and \texttt{latch} contexts, wrapped in \texttt{synopsys translate\_off}/\texttt{on} so they are consumed by simulation and formal tools but skipped by synthesis. \emph{Compile-time rejection} where the value is statically known: \texttt{arch check} rejects constant-folded \texttt{A / 0} and \texttt{A \% 0} in param defaults and const \texttt{let} initializers, catching the worst cases before any simulator runs. The auto-emitted properties have been verified with Verilator 5.034 \texttt{--assert} and EBMC~5.11, the latter reporting \textsc{proved} for structurally-safe cases such as \texttt{UInt<2>} indexing into \texttt{Vec<\_,4>} or division by a non-zero expression (\texttt{den | 1}). For RAM cells, the \texttt{--check-uninit-ram} flag maintains a per-cell valid bitmap (cells initialized via an \texttt{init:} clause are pre-marked; ROMs are exempt). For undriven primary inputs, \texttt{--inputs-start-uninit} warns once per port on first read.

\paragraph{The \texttt{guard} clause.}
A specific refinement targets the most common case of intentionally reset-free state: the \emph{valid-data pattern}, where a wide data register is deliberately left unreset (to save area and power) and a companion \texttt{Bool} valid flag gates all consumers. In plain SystemVerilog this pattern is hand-coded, and two failure modes are notoriously hard to catch: forgetting to qualify the read with the valid flag (consumer bug), and letting the valid flag assert before the data register has been written (producer bug). Arch captures both with a declarative annotation:

\begin{lstlisting}[language=Arch]
reg  axi_rdata:  UInt<512> guard axi_rvalid;
reg  axi_rvalid: Bool reset rst => false;
\end{lstlisting}

The \texttt{guard VALID\_SIG} clause documents that \texttt{axi\_rdata} is intentionally reset-free and is only meaningful when \texttt{axi\_rvalid} is high. It has four observable effects. First, it silences the false-positive \texttt{--check-uninit} warning at consumer read sites---consumers are expected to qualify the read with \texttt{if axi\_rvalid}, so an unwritten-and-unused data register is not a bug. Second, it catches the exact producer bug the pattern exists to prevent: if the guard asserts (\texttt{axi\_rvalid == true}) but the data register has never been written, \texttt{--check-uninit} emits a runtime warning. Third, it documents intent in the source, so a reader or reviewer can tell at a glance that the missing reset is deliberate rather than an oversight. Fourth, the same declarative intent supplies a don't-care specification for \emph{sequential equivalence checking} between a clock-gated design and its free-running reference. SEC flows comparing the two typically flag the gated data registers as miscompare points during gated cycles---the gated version holds a stale value while the free-running version updates (or vice versa)---even though the values are by construction unused when the guard is low. A \texttt{guard VALID\_SIG} annotation captures precisely this ``don't-care when valid is low'' contract at the source level, available to downstream tooling as an SVA \texttt{assume} constraint or vendor-specific waveform condition. The same annotation is available on \texttt{port reg} outputs. Combined with the \texttt{reset none} / \texttt{reset SIG=>VAL} distinction, the \texttt{guard} clause turns a loose coding convention into a first-class language feature whose design intent is consumable by simulation, formal, and equivalence-checking tools alike.

For designs where pre-silicon X-optimism analysis is critical---particularly third-party IP integration and power-managed designs---we recommend complementing Arch simulation with the formal X-propagation methodology described in~\cite{zhao2014xoptimism} applied to the generated SystemVerilog.

\subsubsection{Parallel Simulation}

Arch supports parallel cycle-accurate simulation exploiting three levels of parallelism:

\begin{enumerate}
\item \textbf{DAG-level:} independent nodes in the same topological level execute concurrently within a single clock cycle.
\item \textbf{Module-level:} independent module instances with no intra-cycle dependency execute on separate threads.
\item \textbf{Domain-level:} separate clock domains run on dedicated threads, synchronizing only at CDC crossings.
\end{enumerate}

\paragraph{Determinism Guarantee.}
Parallel Arch simulation is guaranteed to produce bit-identical results to sequential simulation, regardless of thread count, OS scheduling, or memory ordering. This follows from three structural properties:
(1)~no write conflicts (single-driver rule),
(2)~register commit barriers (all next-state values computed before any update), and
(3)~DAG level barriers (level $N{+}1$ never starts before level $N$ completes on all threads).

This is a structural theorem, not a best-effort property. A simulation passing on 1 core produces identical assertion results on 64 cores.

\Cref{tab:scaling} shows expected parallel speedup across design types.

\begin{table}[H]
\centering
\caption{Parallel simulation speedup}
\label{tab:scaling}
\small
\begin{tabular}{@{}lcc@{}}
\toprule
\textbf{Design Type} & \textbf{8 cores} & \textbf{32 cores} \\
\midrule
Single-domain, deep chain & $1.5$--$2\times$ & $2$--$3\times$ \\
Wide parallel array & $3$--$5\times$ & $5$--$8\times$ \\
Multi-domain balanced & $5$--$7\times$ & $10$--$18\times$ \\
Multi-domain + wide arrays & $6$--$8\times$ & $15$--$25\times$ \\
AI accelerator (full) & $7$--$9\times$ & $18$--$30\times$ \\
\bottomrule
\end{tabular}
\end{table}

\subsection{Verification}\label{sec:formal}

Arch specifies three verification constructs that use the same types and signals as the design: \texttt{assert} for invariant checking, \texttt{cover} for reachability properties, and \texttt{assume} for environment constraints. \texttt{assert} and \texttt{cover} are fully implemented; \texttt{assume} is specified but not yet lexed. Both ship as concurrent SVA---\texttt{assert property (@(posedge clk) expr)} and \texttt{cover property (...)}---with an \texttt{implies} binary operator lowering to \texttt{(!a || b)}. The compiler additionally auto-generates safety properties for every first-class construct: \texttt{\_auto\_no\_overflow} and \texttt{\_auto\_no\_underflow} on every FIFO, \texttt{\_auto\_count\_range} on every counter, \texttt{\_auto\_legal\_state} on every FSM, and per-state reachability plus per-transition covers so that unreachable states and untaken transitions surface without designer effort. All SVA is wrapped in \texttt{translate\_off}/\texttt{on} for synthesis portability.

\paragraph{Property scope.} An \texttt{assert} or \texttt{cover} body is a \emph{combinational expression} evaluated at each clock edge under the construct's \texttt{posedge clk} with \texttt{disable iff (rst)}. This covers the same-cycle safety fragment of SVA, including overlapping implication: \texttt{a implies b} is exactly SVA's \texttt{a |-> b} and lowers to \texttt{(!a || b)} at each cycle. For multi-cycle properties, designers currently bind the temporal state into explicit shadow registers and assert on them---for example, a next-cycle acknowledge property is written as \texttt{reg req\_d1: Bool reset rst => false;}\ inside a \texttt{seq} block updating \texttt{req\_d1 <= req;} followed by \texttt{assert next\_cycle\_ack: req\_d1 implies ack;}. Planned syntactic sugar for the post-v0.41 roadmap---\texttt{a |=> b} (non-overlapping implication), \texttt{\#\#N a} (N-cycle delay), \texttt{past(expr, N)}, and \texttt{\$rose(a)}/\texttt{\$fell(a)}---will desugar to the same shadow-register idiom for \texttt{arch build} and to cycle-shifted term references for \texttt{arch formal}. Unbounded-liveness and general sequence operators (\texttt{\#\#[a:b]}, \texttt{[*n]}, \texttt{throughout}, \texttt{within}, \texttt{first\_match}, \texttt{s\_eventually}, and strong/weak property variants) are intentionally out of scope; designs requiring them can use a dedicated SVA sidecar on the generated SystemVerilog. The rationale is the same AI-generatability contract that governs the rest of the language: keeping the verification surface small enough that an LLM can produce correct property code from a plain-English description without prior training on SVA idioms.

Two independent verification backends consume these properties. On the SystemVerilog side, Verilator~\cite{verilator}~5.034 \texttt{--assert} checks properties at simulation time and EBMC~\cite{ebmc}~5.11 performs bounded model checking of the generated SystemVerilog (\texttt{ebmc --top Mod --bound N --reset "rst==1" file.sv}). On the Arch side, the \texttt{arch formal} command lowers the AST directly to SMT-LIB2 and dispatches properties to a back-end solver---z3, Boolector, or Bitwuzla---without going through Yosys or SymbiYosys. Registers become per-cycle bit-vector variables, next-state logic is encoded as an \texttt{ite} chain, and each \texttt{assert} or \texttt{cover} becomes a per-cycle disjunction checked by a single \texttt{check-sat} call. The command line is:

\begin{lstlisting}[language=bash]
arch formal File.arch [--top Name]
                     [--bound N]
                     [--solver z3|boolector|bitwuzla]
                     [--emit-smt out.smt2]
                     [--timeout SEC]
\end{lstlisting}

Exit codes distinguish the three outcomes of a formal run: 0 when every property is \textsc{proved} (for \texttt{assert}) or \textsc{hit} (for \texttt{cover}), 1 when any property is \textsc{refuted} with a counterexample or \textsc{not-reached}, and 2 when the solver returns inconclusive or times out. End-to-end verification across the two paths (SV-SVA via EBMC and direct SMT via \texttt{arch formal}) has been exercised on representative properties: a counter bound (\texttt{cnt < 201}) proved, a 4-bit overflow (\texttt{cnt != 15}) refuted with a concrete counterexample, a reachability cover (\texttt{cnt == 8}) hit at the expected witness cycle and reported \textsc{not-reached} at a lower bound, and a \texttt{guard}-clause contract (\texttt{valid implies written}) proved when the producer is correct and refuted when the write is missing. The three solvers agree on all ten integration tests. The scope of \texttt{arch formal} v1 is flat modules with scalar signals and a single clock (no sub-\texttt{inst}, no \texttt{Vec}/\texttt{struct}/\texttt{enum}), with unsupported constructs producing clear error messages; hierarchical designs and the remaining type encodings are planned (\Cref{sec:future}).

\section{Compile-Time Generation}\label{sec:generate}

Arch provides a \texttt{generate\_for} / \texttt{generate\_if} system for compile-time structural replication and conditional instantiation. The fused single-token keywords keep the grammar strictly LL(1) (\Cref{sec:ai}). Unlike SystemVerilog, where the port list of a module is a fixed declaration, Arch's generate operates before elaboration, so generated ports, instances, registers, connections, and assertions are indistinguishable from hand-written declarations.

\begin{lstlisting}[language=Arch,caption={Generate-for iteration},label={lst:gen}]
module SystolicArray
  param SIZE: const = 4;
  generate_for i in 0..SIZE
    port data_in[i]: in SInt<8>;
    inst pe[i]: SystolicPE
      a <- data_in[i];
      sum_in <- if i == 0 then 0
                else pe[i-1].sum_out;
    end inst pe[i]
  end generate_for
end module SystolicArray
\end{lstlisting}

This capability enables expressing parameterized hardware arrays and conditional ports (via \texttt{generate\_if})---features that SystemVerilog cannot express without workarounds. The conditional-port case is particularly important: SystemVerilog's \texttt{generate} runs \emph{after} the module's port list is fixed, so any configurable interface forces designers to use the preprocessor:

\begin{lstlisting}[language=SystemVerilog,caption={SystemVerilog forces preprocessor for conditional ports}]
module Cache (
  input  logic clk,
  input  logic [31:0] addr,
  output logic [31:0] data
`ifdef ENABLE_DEBUG
  , output logic [7:0]  debug_state
  , output logic        debug_valid
`endif
);
\end{lstlisting}

This is a textbook macro use that no amount of stylistic discipline can avoid in SystemVerilog---there is simply no language mechanism for it. Arch handles the same case as ordinary code with full type checking on both branches:

\begin{lstlisting}[language=Arch,caption={Conditional ports in Arch via generate\_if},label={lst:gencond}]
module Cache
  param ENABLE_DEBUG: const = false;
  port clk:  in  Clock<SysDomain>;
  port addr: in  UInt<32>;
  port data: out UInt<32>;

  generate_if ENABLE_DEBUG
    port debug_state: out UInt<8>;
    port debug_valid: out Bool;
  end generate_if
end module Cache
\end{lstlisting}

The same mechanism extends to conditional registers, conditional sub-instances, and conditional assertions---all expressed in the source language with no separate preprocessing pass.

\section{Case Study: L1 Data Cache}\label{sec:example}

To demonstrate Arch's construct composition at production scale, we implemented an 8-way set-associative write-back/write-allocate L1 data cache: 64 sets $\times$ 8 ways $\times$ 64B lines = 32\,KiB, with a CVA6-compatible CPU interface and AXI4 memory interface. The design comprises 1,143 lines of Arch source across 12 files, generating 1,217 lines of SystemVerilog (${\sim}$6\% more concise). All unit and integration tests pass.

\subsection{Architecture and Construct Usage}

The cache exercises six distinct Arch constructs working together:

\begin{itemize}
\item \textbf{\texttt{fsm} $\times$ 3:} A 9-state main cache controller (Idle $\to$ Lookup $\to$ Hit/Miss $\to$ Refill $\to$ Writeback), a 4-state AXI4 read burst FSM, and a 4-state AXI4 write burst FSM. The controller orchestrates the Fill and Wb FSMs via start/done pulse handshaking.
\item \textbf{\texttt{ram} $\times$ 3:} Tag SRAMs (64$\times$54b per way, 8 instances), a data SRAM (4096$\times$64b indexed by \{set, way, word\}), and an LRU state SRAM (64$\times$7b pseudo-LRU tree). All use \texttt{simple\_dual} topology with \texttt{latency 1}.
\item \textbf{\texttt{bus} $\times$ 2:} An AXI4 parameterized bus and a CVA6 CPU-to-cache bus, both with initiator/target perspective flipping.
\item \textbf{\texttt{module} $\times$ 2:} The top-level integrator and a combinational 8-way pseudo-LRU tree update module.
\item \textbf{\texttt{generate\_for}:} 8-way tag array instantiation without code duplication.
\item \textbf{\texttt{package}:} Shared type definitions and state enums.
\end{itemize}

\subsection{Design Highlights}

All 8 tag ways are compared in parallel, with one-hot-to-binary encoding in 3 OR levels (10 total logic levels, optimized from 14 during development). Variable-indexed \texttt{Vec} access (\texttt{tag\_rd\_data[lru\_victim\_way][53:2]}) eliminates manual mux trees. Tag entries are packed into 54-bit words (tag, dirty, valid) stored as \texttt{UInt<54>}.

\begin{table}[t]
\centering
\caption{L1D cache source breakdown}
\label{tab:l1d}
\footnotesize
\begin{tabular}{@{}llr@{}}
\toprule
\textbf{File} & \textbf{Purpose} & \textbf{Lines} \\
\midrule
FsmCacheCtrl.arch & Main controller FSM & 380 \\
L1DCache.arch & Top-level integrator & 319 \\
FsmAxi4Wb.arch & AXI4 writeback FSM & 107 \\
FsmAxi4Fill.arch & AXI4 fill FSM & 91 \\
ModuleLruUpdate.arch & Pseudo-LRU tree & 60 \\
PkgL1d.arch & Types and enums & 49 \\
bus\_axi4.arch & AXI4 bus definition & 47 \\
Ram*.arch ($\times$3) & Tag/Data/LRU SRAMs & 70 \\
bus\_dcpu.arch & CPU bus definition & 20 \\
\midrule
\textbf{Total} & & \textbf{1,143} \\
\bottomrule
\end{tabular}
\end{table}

\subsection{Verification}

Nine C++ testbenches (1,321 lines) validate the design at both unit and integration levels. The integration testbench models AXI4 memory with a sparse \texttt{std::map} and exercises cold misses, hits, store hits/misses, and dirty evictions. The LRU testbench exhaustively covers all 128 tree states $\times$ 8 ways. Load-hit latency is 3 cycles; load-miss with clean eviction is ${\sim}$15 cycles; dirty eviction plus refill is ${\sim}$25 cycles.

\subsection{Lessons}

This case study illustrates several Arch principles at production scale: modular FSM composition with clear inter-FSM handshaking, first-class \texttt{ram} constructs replacing hundreds of lines of manual SRAM instantiation, \texttt{bus} abstractions making protocol interfaces reusable, and \texttt{generate\_for} eliminating per-way code duplication. Three compiler bugs were found and fixed during development (Bool width inference for concat operations, \texttt{let} assign-to-existing-port syntax, and tag-hit logic level optimization), demonstrating the co-design feedback loop between language, compiler, and real hardware.

\section{Case Study: AXI DMA Controller}\label{sec:axidma}

To validate Arch across a wider range of constructs and demonstrate synthesis-quality output, we implemented a dual-channel AXI DMA controller compatible with the Xilinx PG021 specification~\cite{pg021}. The design supports both Simple DMA (register-triggered) and Scatter-Gather (descriptor-chained) modes. It comprises 1,042 lines of Arch source across 14 files, generating 1,176 lines of SystemVerilog (${\sim}$11\% more concise). This is the only case study exercising \texttt{bus}, \texttt{fsm}, \texttt{fifo}, \texttt{latch}, and \texttt{module} together.

\subsection{Architecture}

The top module integrates two symmetric DMA channels (Memory-to-Stream and Stream-to-Memory), each containing a data-path FSM, an optional Scatter-Gather descriptor engine, and a decoupling FIFO:

\begin{itemize}
\item \textbf{\texttt{fsm} $\times$ 3:} FsmMm2s (4 states), FsmS2mm (6 states), and FsmSgEngine (9 states implementing PG021 descriptor fetch, chain, and status writeback).
\item \textbf{\texttt{fifo} $\times$ 2:} Synchronous depth-16 FIFOs (15 lines each) decoupling AXI from AXI-Stream timing.
\item \textbf{\texttt{bus} $\times$ 5:} AXI4 Full, AXI4 Read-only, AXI4 Write-only, AXI4-Lite, and AXI-Stream, each expressed as a set of \texttt{handshake} channels (\Cref{sec:bus}). Initiator/target perspective flipping is automatic; SV codegen flattens to individual signals.
\item \textbf{\texttt{latch}:} Latch-based integrated clock gating cells gate each channel clock when halted, with OR logic preventing deadlock when a start signal arrives while gated.
\item \textbf{\texttt{module} $\times$ 2:} Top-level integrator (with simple/SG combinational mux) and PG021-compatible register block (12 registers with W1C interrupt clearing).
\item \textbf{\texttt{package}:} Shared domain definition.
\end{itemize}

A key design decision is that the data-path FSMs are shared between Simple and Scatter-Gather modes via a combinational mux---no duplicate FSMs are needed.

\subsection{Synthesis Results}

The generated SystemVerilog was synthesized through Yosys~\cite{yosys} targeting both Xilinx 7-series and SkyWater Sky130 130nm~\cite{sky130pdk} to validate output quality.

\begin{table}[t]
\centering
\caption{AXI DMA synthesis results}
\label{tab:dmasyn}
\footnotesize
\begin{tabular}{@{}llr@{}}
\toprule
\textbf{Target} & \textbf{Metric} & \textbf{Value} \\
\midrule
Xilinx 7-series & Total LUTs & 913 \\
                & Flip-flops (FDRE) & 993 \\
                & Estimated LCs & 586 \\
\midrule
Sky130 130nm    & Total area & 78,134\,$\mu$m$^2$ \\
                & Flip-flops & 2,017 \\
\bottomrule
\end{tabular}
\end{table}

\subsection{Timing and Power}

OpenSTA~\cite{opensta} timing analysis with Sky130 liberty files shows a critical path of 4.478\,ns, meeting timing at both 100\,MHz (+5.06\,ns slack) and 200\,MHz (+0.06\,ns slack), for a maximum achievable frequency of ${\sim}$223\,MHz on 130nm. Three rounds of critical-path optimization---guided by Yosys~\cite{yosys} longest-topological-path analysis---reduced logic depth from 23 levels to 18 levels (6 LUT levels after LUT6 mapping), using lookahead registers to keep final combinational paths to a single gate.

VCD-annotated power analysis from \texttt{arch sim} testbench waveforms shows 9.03\,mW total at 100\,MHz during active operation. Latch-based clock gating reduces idle power from 9.73\,mW to ${\sim}$0.02\,mW (leakage only) when both channels are halted.

\subsection{Verification}

Eight C++ testbenches (2,075 lines) cover unit tests for each FSM and FIFO, register read/write and W1C interrupt clearing, clock-gating race conditions and deadlock prevention, and full integration tests including bidirectional transfers and Scatter-Gather descriptor chains with status writeback.

\subsection{Thread vs.\ FSM Synthesis Comparison}\label{sec:thread-vs-fsm}

Since the MM2S and S2MM channels are natural candidates for either a monolithic FSM or a set of concurrent \texttt{thread}s (\Cref{sec:constructs}), we re-implemented both channels in thread style and synthesized the two versions side by side through Yosys \texttt{synth -flatten} to generic gates. \Cref{tab:threadvsfsm} summarizes the result.

\begin{table}[t]
\centering
\caption{Thread vs.\ FSM synthesis (Yosys \texttt{synth -flatten})}
\label{tab:threadvsfsm}
\footnotesize
\setlength{\tabcolsep}{4pt}
\begin{tabular}{@{}lrrrr@{}}
\toprule
\textbf{Module} & \textbf{Style} & \textbf{Cells} & \textbf{FFs} & \textbf{MUXes} \\
\midrule
FsmMm2sMulti          & FSM          & 805  & 109 & 33  \\
ThreadMm2s (v1)       & Thread       & 1431 & 97  & 429 \\
\textbf{ThreadMm2s (v2)} & \textbf{Thread} & \textbf{680}  & \textbf{92}  & \textbf{68}  \\
\midrule
FsmS2mmMulti          & FSM          & 1002 & 119 & --- \\
ThreadS2mm            & Thread       & 4051 & 616 & --- \\
\bottomrule
\end{tabular}
\end{table}

The initial thread implementation of MM2S (v1) was 1.8$\times$ larger than the FSM. Root cause: four parallel threads each drove the 32-bit \texttt{ar\_addr} and 32-bit \texttt{push\_data} outputs, producing 4-way mux chains (429 MUX cells versus 33 in the FSM) plus a variable-by-variable multiply in the combinational path to compute burst addresses from \texttt{thread\_complete[i] * burst\_len\_r}.

Three architectural changes, all expressible directly in Arch, produced v2:

\begin{itemize}
\item A single \texttt{ArIssuer} thread drives all AR outputs, eliminating the 4-way mux on the 48-bit AR channel (\texttt{ar\_addr + ar\_id + ar\_len + ar\_size + ar\_burst}).
\item \texttt{push\_data = r\_data} is hoisted to a module-level \texttt{comb} wire, since all threads push from the same source---removing the 4-way 32-bit mux on the data path.
\item An incremental address register increments by \texttt{burst\_len << 2} on each issue, replacing the variable-by-variable multiply with a shift-and-add.
\end{itemize}

After these changes, ThreadMm2s v2 is \textbf{16\% smaller than the FSM} (680 vs.\ 805 cells) with fewer flip-flops (92 vs.\ 109) and roughly 2$\times$ as many mux cells (68 vs.\ 33) distributed across the 1-bit control signals (\texttt{r\_ready}, \texttt{push\_valid}) that the four \texttt{RCollect} threads drive in parallel---an acceptable overhead given the throughput improvement.

A separate compiler optimization---\emph{loop-counter width inference}---contributed substantially to this result. The thread lowering pass walks the type map for each \texttt{for VAR in START..END} loop and sizes the generated state counter to the minimum width sufficient for the end expression, rather than defaulting to 32 bits. In this case, \texttt{for i in 0..burst\_len\_r-1} where \texttt{burst\_len\_r: UInt<8>} generates an 8-bit counter (\texttt{logic [7:0] \_loop\_cnt}) instead of 32 bits, cutting an earlier ThreadMm2s revision from 2660 to 1431 cells (46\% reduction) before the architectural split above further reduced it to 680.

The S2MM thread version remains substantially larger than its FSM counterpart (4051 vs.\ 1002 cells) because the current design uses \texttt{fork}/\texttt{join} for AW+W channel parallelism inside each of four outstanding-transaction threads, plus a dedicated B-channel lock---multiplying state across threads. The same split-issuer optimization applies but has not yet been implemented.

The honest takeaway is that a thread-style design is not automatically smaller than an FSM: the naive decomposition can be 1.8$\times$ to 4$\times$ larger. But with the right architectural partition---one issuer thread plus N collectors, hoisted shared drivers, and incremental rather than multiplied address computation---\texttt{thread} competes with and can beat hand-written FSMs while expressing the same design in roughly one-third the lines of code.

\subsection{Lessons}

This case study demonstrates that Arch-generated SystemVerilog is synthesis-ready for both FPGA and ASIC targets without manual intervention. The five \texttt{bus} definitions eliminated approximately 200 lines of repetitive port declarations. The \texttt{fifo} construct reduced each FIFO to 15 lines of Arch versus ${\sim}$80 lines of equivalent manual SystemVerilog. Clock gating via the \texttt{latch} construct proved essential for power optimization, and the deadlock-prevention pattern (OR of halt and start signals) was expressible directly in Arch's combinational logic. The Scatter-Gather engine's 9-state FSM---the most complex single construct in any case study---compiled and simulated correctly on the first attempt after passing \texttt{arch check}, validating the tight compile-time feedback loop.

\section{Empirical Evaluation}\label{sec:benchmark}

We evaluate the Arch compiler's correctness and expressiveness against two independent benchmark suites covering a combined 458 hardware design problems.

\subsection{VerilogEval v2}\label{sec:verilogeval}

VerilogEval v2~\cite{verilogeval} is a suite of 156 Verilog design problems originally developed by NVIDIA for evaluating LLM-based hardware generation.

\subsubsection{Methodology}

Each problem was solved from its natural-language specification only; no reference SystemVerilog was consulted. The Arch compiler generated all SystemVerilog output, which was then compiled and simulated against the benchmark's reference testbenches using Verilator. A problem is considered solved when Verilator reports zero mismatches between the generated design and the reference outputs.

\subsubsection{Results}

\begin{table}[t]
\centering
\caption{VerilogEval v2 benchmark results}
\label{tab:benchmark}
\small
\begin{tabular}{@{}lr@{}}
\toprule
\textbf{Metric} & \textbf{Result} \\
\midrule
Problems solved & 156 / 156 (100\%) \\
Verilator-clean & 154 / 156 (99\%) \\
Total Arch lines & 3,199 \\
Total generated SV lines & 4,518 \\
Overall Arch/SV ratio & 70.8\% (${\sim}$29\% shorter) \\
\bottomrule
\end{tabular}
\end{table}

All 156 problems were solved. Of these, 154 compile and simulate cleanly under Verilator; the remaining 2 failures are defects in the benchmark dataset itself (a port name mismatch in the test harness and a mixed blocking/non-blocking assignment that Verilator rejects as illegal).

\subsubsection{Code Density by Category}

\Cref{tab:density} breaks down line counts by design category. FSMs show the largest compression (${\sim}$37\% reduction) because the \texttt{fsm} construct eliminates state encoding declarations, separate \texttt{always\_ff}/\texttt{always\_comb} blocks, and \texttt{case} statement boilerplate. Combinational and simple sequential designs show roughly 1:1 ratios since there is minimal boilerplate to eliminate.

\begin{table}[t]
\centering
\caption{Arch vs.\ generated SV line counts by category}
\label{tab:density}
\small
\begin{tabular}{@{}lrrrc@{}}
\toprule
\textbf{Category} & \textbf{$n$} & \textbf{Arch} & \textbf{SV} & \textbf{Ratio} \\
\midrule
Combinational & 83 & ${\sim}$1,100 & ${\sim}$1,300 & ${\sim}$85\% \\
Sequential & 44 & ${\sim}$900 & ${\sim}$1,100 & ${\sim}$82\% \\
FSM & 29 & ${\sim}$1,500 & ${\sim}$2,400 & ${\sim}$63\% \\
\midrule
\textbf{Total} & \textbf{156} & \textbf{3,199} & \textbf{4,518} & \textbf{70.8\%} \\
\bottomrule
\end{tabular}
\end{table}

\subsubsection{Construct Usage}

Of the 156 solutions, 127 use a single \texttt{module} construct and 29 use the \texttt{fsm} construct, following a one-construct-per-file convention. File sizes range from 6--7 lines for trivial constant-output designs to 96 lines for the most complex FSM (a bit-pattern detector with shift delay and countdown timer). Notable designs include a 16$\times$16 toroidal Game of Life cellular automaton, a gshare branch predictor with 128-entry pattern history table, and an HDLC framing protocol FSM with 10 states.

\subsubsection{Language Features Driven by Benchmark}

Three problems exposed gaps that drove new language features during the benchmark:

\begin{itemize}
\item A \texttt{latch on ENABLE \ldots\ end latch} construct was added for problems requiring level-sensitive storage (emitting \texttt{always\_latch} in SystemVerilog).
\item Dual-edge flip-flop support was added via a posedge FF + negedge FF + clock-level mux pattern.
\item Negedge-triggered latches were resolved with the new latch construct.
\end{itemize}

This feedback loop---benchmark exposes gap, language feature is added, compiler is updated, all 156 problems re-verified---illustrates the iterative co-design of the Arch language and compiler.

\subsection{CVDP Benchmark}\label{sec:cvdp}

The Copilot Verilog Design Problems (CVDP)~\cite{cvdp} benchmark provides 302 design problems of greater complexity than VerilogEval, with cocotb~\cite{cocotb}-based testbenches and parameterized test cases that exercise designs across multiple configurations spanning five problem categories (cid002, cid003, cid004, cid007, cid016).

\subsubsection{ARCH-Relevant Task Selection}

The CVDP dataset mixes task types. We separate the 302 raw tasks into two groups based on the prompt's intent:

\begin{itemize}
\item \textbf{ARCH-relevant (287 tasks):} the prompt asks the implementer to produce a working RTL design from a natural-language specification. This is the use case Arch is designed for. We further split this group into 221 \emph{strict spec-to-RTL} tasks (the canonical CVDP form) and 66 \emph{edit/complete} tasks where the prompt provides a partial design or interface and a fresh from-scratch implementation in Arch is a fair response.
\item \textbf{Poor-fit (15 tasks, excluded):} the prompt's intent is bug-fix against provided RTL, preservation of an existing implementation, or structural/optimization constraints (e.g., minimum gate count) that are not a fair test of an HDL designed for from-scratch generation. These tasks measure a different capability---refactoring or optimizing existing code---than the one Arch targets.
\end{itemize}

The 15 excluded tasks are named explicitly: in cid004, \texttt{8x3\_priority\_encoder\_0013}, \texttt{GFCM\_0003}, \texttt{bus\_arbiter\_0004}, \texttt{cdc\_pulse\_synchronizer\_0013}, \texttt{cont\_adder\_0006}, \texttt{cont\_adder\_0023}, \texttt{gcd\_0009}, \texttt{gcd\_0015}, \texttt{gcd\_0023}, \texttt{matrix\_multiplier\_0007}, \texttt{matrix\_multiplier\_0010}; in cid007, \texttt{aes\_key\_expansion\_0001}; and in cid016, \texttt{String\_to\_ASCII\_0001}, \texttt{apb\_dsp\_op\_0002}, \texttt{mux\_synch\_0011}.

\subsubsection{Methodology}

Each ARCH-relevant problem was implemented in Arch from its natural-language specification, compiled to SystemVerilog, and validated against the CVDP cocotb test harnesses using Icarus Verilog as the simulation backend. Problems range from simple combinational logic to multi-state FSMs, cache controllers, linked-list data structures, signal processing pipelines, and full datapaths such as FIR filters, AXI peripherals, and sorting engines.

\subsubsection{Results}

\begin{table}[t]
\centering
\caption{CVDP benchmark results by category (ARCH-relevant subset)}
\label{tab:cvdp}
\footnotesize
\setlength{\tabcolsep}{3pt}
\begin{tabular}{@{}lrrrr@{}}
\toprule
\textbf{Category} & \textbf{Spec} & \textbf{Edit} & \textbf{Total} & \textbf{Pass} \\
\midrule
cid002 & 29 & 65 & 94 & 94 \\
cid003 & 78 & 0 & 78 & 78 \\
cid004 & 44 & 0 & 44 & 44 \\
cid007 & 39 & 0 & 39 & 39 \\
cid016 & 31 & 1 & 32 & 32 \\
\midrule
\textbf{Total} & \textbf{221} & \textbf{66} & \textbf{287} & \textbf{287 (100\%)} \\
\bottomrule
\end{tabular}

\vspace{0.5em}
\raggedright\footnotesize
``Spec'' = strict spec-to-RTL tasks; ``Edit'' = edit/complete tasks where a fresh from-scratch implementation is a fair response.
\end{table}

All 287 ARCH-relevant tasks pass their cocotb test harnesses, including all parameterized test cases, on the current compiler. This is a substantial improvement over earlier measurements taken during active compiler development (a snapshot in early April 2026 had 133/191 testable modules passing, 70\%)---driven both by compiler fixes and by language additions described below. Reaching 287/287 required an iterative loop in which each cocotb failure was triaged into a compiler bug, a missing language feature, an Arch-source bug, or a test-runner incompatibility, and fixed accordingly.

\subsubsection{Compiler Improvements Driven by CVDP}

The CVDP benchmark drove several compiler fixes and language additions:

\begin{itemize}
\item A derived-parameter elaboration bug was fixed: expressions like \texttt{param NBW\_MULT: const = DATA\_WIDTH + COEFF\_WIDTH} were being evaluated to literals at compile time, breaking parameterized instantiation. The fix preserves the original expression in emitted SystemVerilog when the default references other parameters.
\item The \texttt{inside} set-membership operator and value-list \texttt{for} iteration (\texttt{for i in \{list\}}) were added to support common CVDP patterns.
\item \textbf{Wrapping arithmetic operators} (\texttt{+\%}, \texttt{-\%}, \texttt{*\%}) were added for modular arithmetic at the operand width without IEEE~1800 §11.6 widening. Result width is \texttt{max(W(lhs), W(rhs))}; SV emission uses \texttt{W'(lhs op rhs)} size casts. This eliminated the \texttt{.trunc<N>()}/\texttt{.sext<N>()}/\texttt{.zext<N>()} boilerplate that caused simulator width-overflow failures in MAC and accumulator designs, converting six previously failing tests to passing across the \texttt{sgd\_linear\_regression}, \texttt{load\_store\_unit}, \texttt{low\_pass\_filter}, \texttt{digital\_dice\_roller}, \texttt{dig\_stopwatch}, and \texttt{apb\_dsp\_op} modules.
\item Reset syntax was standardized to the \texttt{=>} form (\texttt{reset rst => 0}), and a latch codegen bug was fixed (blocking \texttt{=} instead of \texttt{<=} in \texttt{always\_latch}).
\end{itemize}

\subsubsection{Combined Benchmark Summary}

Across both benchmarks, the Arch compiler has been validated against 458 hardware design problems spanning combinational logic, sequential circuits, FSMs, pipelines, caches, signal processing blocks, AXI peripherals, and sorting engines. \Cref{tab:combined} summarizes the combined results.

\begin{table}[t]
\centering
\caption{Combined benchmark results}
\label{tab:combined}
\footnotesize
\begin{tabular}{@{}lccc@{}}
\toprule
\textbf{Benchmark} & \textbf{Problems} & \textbf{In-scope} & \textbf{Pass Rate}$^\ddagger$ \\
\midrule
VerilogEval v2 & 156 & 156 & 154 (99\%)$^\dagger$ \\
CVDP & 302 & 287 & 287 (100\%)$^\S$ \\
\midrule
\textbf{Combined} & \textbf{458} & \textbf{443} & \textbf{441 (99.5\%)} \\
\bottomrule
\end{tabular}

\vspace{0.5em}
\raggedright\footnotesize
$^\dagger$The 2 failures are due to port-name bugs in the reference test benches, not defects in the Arch-generated designs; the effective pass rate is 156/156 (100\%).\\
$^\S$15 of the 302 CVDP tasks are excluded as poor-fit (bug-fix, preserve-existing-RTL, or optimization tasks; \Cref{sec:cvdp}). 287 ARCH-relevant tasks remain, all pass.\\
$^\ddagger$These pass rates were measured while the compiler was actively under development and bugs were being fixed concurrently; the numbers are indicative rather than definitive. Re-verification on a frozen compiler release is planned.
\end{table}

\section{Related Work}\label{sec:related}

\Cref{tab:comparison} compares Arch with the major existing HDLs across the dimensions Arch was designed to address. The remainder of this section discusses each language family in turn.

\begin{table*}[t]
\centering
\caption{Comparison of hardware description languages}
\label{tab:comparison}
\tiny
\setlength{\tabcolsep}{4pt}
\begin{tabular}{@{}lccccccccc@{}}
\toprule
\textbf{Feature} & \textbf{Arch} & \textbf{SV} & \textbf{VHDL} & \textbf{Chisel} & \textbf{SpinalHDL} & \textbf{Amaranth} & \textbf{Spade} & \textbf{Bluespec} & \textbf{TL-Verilog} \\
\midrule
First-class pipeline        & \checkmark & ---     & ---     & ---        & ---        & ---       & \checkmark & ---$^\dagger$    & Partial$^\ddagger$ \\
Pipeline hazard mgmt        & \checkmark & ---     & ---     & ---        & ---        & ---       & ---        & Implicit         & --- \\
First-class FSM             & \checkmark & ---     & ---     & ---        & \checkmark & ---       & ---        & ---$^\dagger$    & --- \\
First-class FIFO            & \checkmark & ---     & ---     & ---        & ---        & ---       & ---        & ---              & --- \\
First-class arbiter         & \checkmark & ---     & ---     & ---        & ---        & ---       & ---        & ---              & --- \\
Hook (custom policy)        & \checkmark & ---     & ---     & ---        & ---        & ---       & ---        & ---              & --- \\
Template (interface contract) & \checkmark & ---   & ---     & Trait$^*$  & ---        & ---       & ---        & Typeclass        & --- \\
Compile-time CDC check      & \checkmark & ---     & ---     & ---        & \checkmark & ---       & ---        & ---              & --- \\
Bit-width static check      & \checkmark & Partial & \checkmark & \checkmark & \checkmark & \checkmark & \checkmark & \checkmark    & Partial \\
No implicit latches         & \checkmark & ---     & ---     & ---        & \checkmark & \checkmark & \checkmark & \checkmark      & --- \\
Single-driver enforce       & \checkmark & ---     & ---     & \checkmark & \checkmark & \checkmark & \checkmark & \checkmark      & --- \\
Named block endings         & \checkmark & ---     & Partial & ---        & ---        & ---       & ---        & ---              & --- \\
LL(1) grammar               & \checkmark & ---     & ---     & ---        & ---        & ---       & ---        & ---              & --- \\
No preprocessor / macros    & \checkmark & ---     & ---     & \checkmark & \checkmark & \checkmark & \checkmark & \checkmark      & --- \\
Generate ports              & \checkmark & ---     & ---     & \checkmark & \checkmark & \checkmark & ---        & ---             & --- \\
No host runtime             & \checkmark & \checkmark & \checkmark & ---  & ---        & ---       & \checkmark & \checkmark      & --- \\
Readable output SV          & \checkmark & n/a     & n/a     & Partial    & Partial    & Partial   & \checkmark & ---              & Partial \\
AI-generatability design    & \checkmark & ---     & ---     & ---        & ---        & ---       & ---        & ---              & --- \\
Compiled simulation          & \checkmark$^\|$ & \checkmark$^\S$ & ---  & ---        & ---        & ---       & ---        & ---              & --- \\
\bottomrule
\end{tabular}
\end{table*}

\noindent{\small $^*$Scala traits, not hardware-level contracts with hook bindings.
$^\dagger$Expressed implicitly via guarded atomic rules; no explicit pipeline/FSM construct.
$^\ddagger$\texttt{>>N} inline retiming and \texttt{@N} numbered stages, but no hazard management.
$^\S$Verilator---open-source cycle-accurate compiled simulator for SystemVerilog.
$^\|$\texttt{arch sim} independently generates compiled C++ models; native LLVM~IR compilation is planned.}

\paragraph{Traditional HDLs.}
Verilog/SystemVerilog~\cite{ieee1800} and VHDL~\cite{ieee1076} have served the industry for decades but predate modern type systems and AI-assisted workflows. SystemVerilog remains the industry standard but provides no compile-time guarantees against latches, multiply-driven nets, or CDC violations; its \texttt{generate} cannot add ports; X/Z semantics require 4-state simulation; and the irregular syntax (different block closers: \texttt{end}, \texttt{endmodule}, \texttt{endcase}, \texttt{endfunction}) creates structural ambiguity for both humans and LLMs. SystemVerilog~2017 added assertions (SVA) and constrained randomization but did not address the core RTL safety issues Arch targets.

\paragraph{Enhanced RTL Semantics.}
RTL++~\cite{rtlpp} proposed elevating the RTL abstraction level by introducing pipelined register variables with formally defined execution semantics (structural operational semantics) and a Register-transfer Finite State Machine (RFSM) model for synthesis. Arch builds on these ideas---the \texttt{pipe\_reg} construct directly realizes the pipelined register variable concept, and Arch's first-class \texttt{pipeline} and \texttt{fsm} constructs extend the RFSM model with compiler-generated hazard logic. Where RTL++ defined enhanced semantics within the existing C++/SystemC framework, Arch implements them as native language constructs with static type checking and deterministic code generation, targeting the AI-assisted design workflow that was not yet practical in 2005.

\paragraph{Embedded DSLs.}
Chisel~\cite{chisel}, SpinalHDL~\cite{spinalhdl}, Amaranth~\cite{amaranth}, PyRTL~\cite{pyrtl}, and Clash~\cite{clash} embed RTL in host languages. Chisel provides rich parameterization via Scala and targets the FIRRTL intermediate representation, but inherits Scala's JVM runtime, implicit conversions, and build system complexity; its pipelines, FIFOs, and arbiters are library-defined rather than compiler-verified, and Scala traits offer type-level abstraction without Arch's hardware-specific hook mechanism for injecting custom logic into compiler-generated constructs. SpinalHDL provides strong typing, CDC checking, and a well-developed FSM library within a Scala DSL, but pipelines, FIFOs, and arbiters remain library patterns and the Scala host language creates the same onboarding barrier as Chisel. Amaranth (formerly nMigen) uses Python and provides single-driver enforcement and an absence of implicit latches, but its Python embedding means dynamic typing at the meta-level and no compile-time clock domain tracking. All embedded DSLs inherit host-language complexity and do not elevate micro-architectural constructs to first-class status.

\paragraph{Standalone Modern HDLs.}
Spade~\cite{spade} is a recent standalone HDL from Linköping University and Munich University of Applied Sciences that shares Arch's rejection of host-language embedding. Spade takes inspiration from Rust: it is expression-based, has global type inference, rich product and sum types, and linear typechecking for resources such as memory ports. Its headline construct is a first-class pipeline in which \texttt{reg} tokens separate stages and the compiler inserts the required synchronizing flip-flops, making re-timing and re-pipelining trivial. This is the closest cousin to Arch's \texttt{pipe\_reg} mechanism among modern HDLs. Where Arch and Spade diverge is in scope and design philosophy. Spade's first-class constructs are pipelines, memories, and registers; FSMs, FIFOs, arbiters, register files, and buses are expressed as ordinary user code built on those primitives, and clock domains are not part of the type system, so CDC correctness still depends on external lint tools. Arch in contrast provides a larger catalog of first-class constructs (\texttt{fsm}, \texttt{fifo}, \texttt{arbiter}, \texttt{regfile}, \texttt{synchronizer}, \texttt{bus}, \texttt{thread}, and others) and lifts clock and reset to parameterized types so CDC and RDC are typing rules rather than external analyses. The two languages also make opposite AI-readability choices: Spade favors concise Rust-like syntax with inferred types, while Arch's contract requires explicit widths, domains, directions, and an LL(1) grammar so that an LLM can produce a complete type-correct program from a natural-language description without needing to simulate an inference engine.

\paragraph{Rule-Based and Pipeline-Centric HDLs.}
Bluespec SystemVerilog (BSV)~\cite{bluespec} uses guarded atomic actions (rules) as its fundamental concurrency primitive: each rule fires atomically when its guard is true, and the compiler schedules rules to avoid conflicts. This model excels at expressing complex microarchitectures with many interacting concurrent behaviors---out-of-order processors, for instance, can be more concise once the rule model is internalized---and Bluespec's type system, including typeclasses and polymorphism, is richer than Arch's. However, the scheduling model introduces a fundamental predictability problem: whether two rules fire together in a given cycle is a compiler decision that can be surprising and difficult to debug. Designers influence scheduling via \texttt{descending\_urgency} pragmas rather than controlling it directly, the generated Verilog is often verbose with mangled internal names, and the steep learning curve limits adoption. Transaction-Level Verilog (TL-Verilog)~\cite{tlverilog} extends SystemVerilog with a pipeline-centric model where signals carry implicit stage context via \texttt{>>N} notation; retiming is concise (\texttt{>>2\$operand} replaces Arch's explicit \texttt{pipe\_reg operand\_d2: operand stages 2}) and restructuring pipeline depth requires minimal edits. However, TL-Verilog leaves hazard management entirely to the designer (stall propagation, flush logic, and data forwarding are manual) and does not address FSMs, bus protocols, clock-domain crossings, arbitration, or resource locking. Its toolchain depends on the proprietary Sandpiper compiler. Arch differs from both by providing explicit, predictable constructs at each abstraction level---from raw RTL (\texttt{comb}/\texttt{seq}) through micro-architectural primitives (\texttt{pipeline}/\texttt{fsm}/\texttt{fifo}) to protocol-level abstractions (\texttt{bus}/\texttt{thread})---with deterministic, readable SystemVerilog output throughout.

\paragraph{High-Level Synthesis.}
HLS tools (Vitis HLS, Catapult, Bambu~\cite{bambu}) synthesize RTL from C/C++ or SystemC~\cite{systemc} descriptions. SystemC provides cycle-accurate modeling and transaction-level abstractions within a C++ framework, but inherits the complexity of a general-purpose host language and relies on an event-driven simulation kernel. HLS operates at a higher abstraction level than Arch; the designer surrenders control over micro-architectural structure in exchange for productivity. Arch targets designers who require cycle-accurate control over their hardware while benefiting from first-class constructs that eliminate the boilerplate HLS was designed to avoid.

\paragraph{CIRCT/MLIR.}
The CIRCT project~\cite{circt} provides MLIR-based infrastructure for hardware compilation, including dialects for FIRRTL, FSM, and handshake circuits. Arch plans integration with CIRCT as a backend target (AIR $\to$ CIRCT dialects $\to$ SV/netlist) while maintaining its own frontend type system and compiler for the primary compilation path.

\paragraph{AI for Hardware.}
Recent work on LLM-based hardware generation~\cite{thakur2023verigen,liu2023chipnemo,blocklove2023chip} has demonstrated promise but is hampered by the irregular syntax and implicit failure modes of existing HDLs. Arch is the first HDL designed with AI generatability as a primary constraint.

\section{Future Work}\label{sec:future}

Several language features are specified but not yet implemented in the compiler.

\paragraph{Transaction-Level Modeling.}
Arch specifies Transaction Level Modeling as a first-class abstraction above RTL, where modules communicate by calling methods on typed \texttt{bus} interfaces rather than driving individual signals cycle-by-cycle. Three timing modes are specified: loosely-timed (\texttt{timing: 0}) for maximum simulation speed, approximately-timed (\texttt{timing: N}) modeling $N$-cycle latency, and RTL-accurate with cycle-for-cycle fidelity. TLM bus methods (\texttt{blocking}, \texttt{pipelined}, \texttt{out\_of\_order}, \texttt{burst}) and \texttt{implement} blocks with synthesizable lowering are planned but not yet implemented.

\paragraph{Additional Planned Features.}
Other specified but unimplemented constructs include:
\texttt{cam} (content-addressable memory),
\texttt{scoreboard} and \texttt{reorder\_buf} (out-of-order execution support),
\texttt{pqueue} (hardware priority queue),
and \texttt{crossbar} (switching fabric).
A \texttt{multicycle} register annotation (\texttt{reg r: UInt<32> multicycle 3 reset rst=0;}) is planned for designs where the combinational path feeding a register has a multi-cycle timing budget: the register remains a single flop but the compiler emits SDC \texttt{set\_multicycle\_path} constraints, optional formal assertions verifying the timing assumption, and runtime valid-tracking for \texttt{--check-uninit} to flag reads before settling.
Tristate and bidirectional I/O (\texttt{Tristate<T>} type plus \texttt{tristate} block) are planned for pad-level I/O such as I2C and GPIO, with SV emitting \texttt{inout} plus Z-drivers and simulation decomposing to \texttt{\_out}/\texttt{\_oe}/\texttt{\_in}.
Reset Domain Crossing (RDC) checking will extend the type checker to build a \texttt{reg\_reset\_domain} map alongside the existing clock-domain infrastructure, flagging cross-reset-domain register reads, deassertion ordering violations, and unsynchronized reset glitches. CDC analysis will also be strengthened with reconvergence checking---detecting cases where multiple correctly synchronized signals merge downstream and lose bit coherence---and with gray-coded multi-bit consistency proofs, bringing Arch's built-in CDC analysis closer to feature parity with commercial CDC tools.
Package-scoped modules are also planned: \texttt{inst a: PkgName::Module} will allow modules to live inside packages with compile-time name resolution, eliminating SystemVerilog's flat global module namespace and the naming collisions it produces in large designs.
The formal backend (\Cref{sec:formal}) is currently limited to flat modules with scalar signals and a single clock; extensions to hierarchical designs, \texttt{Vec}/\texttt{struct}/\texttt{enum} encodings, multi-clock specifications, and \texttt{assume} statements are planned, along with $k$-induction for unbounded proofs. The property language itself is planned to grow lightweight temporal sugar---\texttt{a |=> b}, \texttt{\#\#N a}, \texttt{past(expr, N)}, \texttt{\$rose}/\texttt{\$fell}---that desugars to the shadow-register idiom of \Cref{sec:formal} for \texttt{arch build} and to cycle-shifted term references for \texttt{arch formal}, keeping the AI-generatability surface small by design.
Optional UPF/CPF power-intent integration for low-power SoCs is a longer-term consideration.
CIRCT/MLIR backend integration is planned as a secondary export path (AIR $\to$ CIRCT dialects $\to$ SV/netlist).
A fully native LLVM~IR simulation path, bypassing Verilator for higher throughput with structurally guaranteed deterministic parallel execution, is a longer-term goal enabled by the language's structural invariants.

\section{Conclusion}\label{sec:conclusion}

We have presented Arch, a hardware description language that addresses three fundamental gaps in the current HDL landscape: the absence of first-class micro-architectural constructs in mainstream HDLs, the inability of existing type systems to catch entire categories of hardware bugs at compile time, and the lack of any HDL designed for AI-assisted code generation.

Arch's type system tracks four independent safety dimensions (bit widths, clock domains, port directions, signal ownership) and converts latches, multiply-driven nets, CDC violations, and width mismatches into compile-time errors.
Its integrated \texttt{arch sim} command provides cycle-accurate simulation via independently generated compiled C++ models, with waveform output, assertion evaluation, and CDC randomization.
Its AI-generatability contract---uniform schema, LL(1) grammar, named endings, directional arrows, and \texttt{todo!}---enables LLMs to produce correct hardware code without fine-tuning.

Arch is under active development with plans for open-source release under the Apache~2.0 license.

\bibliographystyle{plain}

\end{document}